\documentclass[dvips,apj]{emulateapj}
\usepackage{apjfonts,ifthen}

\newcommand{\asca}{\textit{ASCA}}
\newcommand{\xmm}{\textit{XMM}}

\newcommand{\snt}{SN\,II}
\newcommand{\sntx}{SN\,IIx}
\newcommand{\sno}{SN\,Ia}
\newcommand{\msun}{M$_\sun$}

\submitted{Submitted to the Astrophysical Journal, 2 May 2003}

\shorttitle{GALAXY CLUSTER ELEMENTAL ABUNDANCES}
\shortauthors{BAUMGARTNER ET AL.}

\begin{document}

\title{Intermediate Element Abundances in Galaxy Clusters}

\author{W. H. Baumgartner\altaffilmark{1,2,3}}
\author{M. Loewenstein\altaffilmark{2,1}}
\author{D. J. Horner\altaffilmark{4}}
\author{R. F. Mushotzky\altaffilmark{2}}

\altaffiltext{1}{Astronomy Department, University of Maryland,
	College Park, MD 20742}
\altaffiltext{2}{Laboratory for High Energy Astrophysics, NASA/GSFC, 
	Code 662, Greenbelt, MD 20771}
\altaffiltext{3}{Email: {\tt wayne@astro.umd.edu}}
\altaffiltext{4}{Astronomy Department, University of Massachusetts,
	Amherst, MA 01003}

\begin{abstract}
We present the average abundances of the intermediate elements
obtained by performing a stacked analysis of all the galaxy clusters
in the archive of the X-ray telescope \asca.  We determine the
abundances of Fe, Si, S, and Ni as a function of cluster temperature
(mass) from 1 -- 10\,keV, and place strong upper limits on the
abundances of Ca and Ar.  In general, Si and Ni are overabundant with
respect to Fe, while Ar and Ca are very underabundant.  The
discrepancy between the abundances of Si, S, Ar, and Ca indicate that
the $\alpha$-elements do not behave homogeneously as a single group.
We show that the abundances of the most well-determined elements Fe,
Si, and S in conjunction with recent theoretical supernovae yields do
not give a consistent solution for the fraction of material produced
by Type~Ia and Type~II supernovae at any temperature or mass.  The
general trend is for higher temperature clusters to have more of their
metals produced in Type~II supernovae than in Type~Ias.  The
inconsistency of our results with abundances in the Milky Way indicate
that spiral galaxies are not the dominant metal contributors to the
intracluster medium (ICM).  The pattern of elemental abundances
requires an additional source of metals beyond standard \sno\ and
\snt\ enrichment.  The properties of this new source are well matched
to those of Type~II supernovae with very massive, metal-poor
progenitor stars.  These results are consistent with a significant
fraction of the ICM metals produced by an early generation of
population III stars.

\end{abstract}


\section{INTRODUCTION}

Galaxy clusters provide an excellent environment for determining the
relative abundances of the elements.  Because clusters are the largest
potential wells known, they retain all the enriched material produced
by the member galaxies.  This behavior is in stark contrast to our own
Milky Way \citep{tww95} and many other individual galaxies
\citep{hw99}.  The accumulation of enriched material in clusters can
be used as a probe to study the star formation history of the
universe, the mechanisms that eject the elements into the ICM, the
relative importance of different classes of supernovae, and ultimately
the source of the metals in the intra-cluster medium (ICM).

The dominant baryonic component in clusters is the hot gas in the ICM,
with 5--10 times as much mass as resides in the stellar component.
The physics describing the dominant emission mechanism of the ICM gas
is relatively simple.  The ICM is optically thin, well modeled by a
sphere of hydrostatic gas in thermal equilibrium, and the high
temperatures and moderate densities minimize the importance of dust.
Extinction, ionization, non-equilibrium and optical depth effects are
minimal.  As a result, cluster abundance determinations are more
physically robust and reliable than those in, e.g., stellar systems,
\ion{H}{2} regions, and planetary nebulae.  The hot gas emits
dominantly by thermal bremsstrahlung in the X-ray band, and the strong
transitions to the n=1 level (K-shell) and to the n=2 level (L-shell)
of the H-like and He-like ions of the elements from carbon to nickel
also lie in the X-ray band.  This makes the X-ray band an attractive
place for elemental abundance determinations.

Early X-ray observations of galaxy clusters \citep{mitch76,s77} showed
that the strong H- and He-like iron lines at 6.9 and 6.7\,keV could
lead to a value for the metal abundance in clusters.  Later results
\citep{mush78, mush83} derived from iron line observations showed that
clusters had metal abundances of about $1/3$ the solar value.

The improved spectral resolution and large collecting area of the
\asca\ X-ray telescope brought new power to studies of cluster metal
abundances. In particular, the improved 0.5--10.0\,keV energy range of
\asca\ allowed for better spectroscopic fits to clusters than was
possible with \textsl{ROSAT}, which had an upper limit of 2.5\,keV.
\citet{ml96} studied four bright clusters at temperatures such that
strong line emission is present, and provided the first measurements
of elemental abundances other than iron since the initial results from
{\it Einstein} \citep{m81,b79,r84}.  Their measurements of silicon,
neon and sulfur were interpreted as high abundances of the
$\alpha$-elements in clusters.  This result suggested that type II
supernovae (which produce much higher $\alpha$ element yields than
\sno) from massive stars are responsible for a significant fraction of
the metals in the ICM.  (Type Ia supernovae (\sno) produce high yields
of elements in the iron peak, while Type II supernovae (\snt) produce
yields rich in the $\alpha$ elements Si, S, Ne, and Mg.)  Later work
by \cite{fuk97} showed that clusters are more metal enriched in their
centers, and that the Si/Fe ratio is about 1.5--2.0 with respect to
the solar value.  \citet{fuk98} showed how the silicon abundance was
higher in hotter clusters, and confirmed the importance of \snt\ in
cluster enrichment.  More recently, \citet{f00}, \citet{f01}, and
\citet{f02} used \asca\ and \xmm\ data to show that type Ia products
dominate in the centers of certain clusters and how type II products
are more evenly distributed.  The observation by \cite{a92} that the
metal mass in clusters is correlated with the optical light from early
type galaxies and not from spirals is also important in determining
the origins of metals in the ICM.

In this paper we use the \asca\ satellite \citep{tih} to further
constrain the abundances of the intermediate elements.  Previous
authors referred to the elements Ne, Mg, Si, S, Ca and Ar as
$\alpha$-elements in order to emphasize their supposed similar
formation mechanism; we will refer to the elements observable with
X-ray spectroscopy in the \asca\ band as intermediate elements.  This
label includes nickel in the group and is preferred since the
observations will show that \emph{the $\alpha$-elements do not act
homogeneously as a single class}.

The large database of over 300 cluster observations makes the \asca\
satellite well suited for a overall analysis of the intermediate
element abundances in galaxy clusters.  While it is not possible to
obtain accurate abundances of these elements for more than a few
individual clusters, we jointly analyze many clusters at a time in
several ``stacks'' in order to obtain the signal necessary for
obtaining the abundances.  The relatively large field of view of the
\asca\ telescope allows for spectroscopic analysis of the entire
spatial extent of all but the closest clusters, and the moderate
spectral resolution enables abundance determinations from the K-shell
and L-shell lines.

\cite{fuk97} and other observations of clusters obtained with
\textsl{Chandra} and \xmm\ have show that abundance gradients are
common across the spatial extent of clusters, often with enhanced iron
abundances in the cluster centers.  These observations shed valuable
light on the source of the metals and help discriminate among the
mechanisms that enrich the ICM.  \cite{dig}\footnote{The proceedings
of the Ringberg Cluster Conference, \citep{dig} can be found at: {\tt
http://www.xray.mpe.mpg.de/\~{}ringberg03/}} has shown with {\it
BeppoSax} measurements that the centers of clusters (within a radius
where the density is 3500 times the critical density) have iron
abundances that are enhanced by about 10--20\%.  These results
indicate the importance of a physical mechanism in the very center of
clusters that causes an increase in the central metallicity.  However,
this occurs only at small radii and does not influence average
abundance measurements integrated out to large radii where most of the
cluster mass resides.

With these cluster elemental abundances, we investigate the source of
the metals as a mixture of canonical \sno\ and \snt, and propose
alternative sources of metals necessary to match the observations.

\section{THE ELEMENTS}

The strong n=2 to n=1 Ly-$\alpha$ (or K-$\alpha$) lines of the
elements from C to Ni lie in the X-ray band between 0.1--10.0\,keV.
These are the largest equivalent width lines in the X-ray spectrum for
clusters with temperatures greater than $\sim$2\,keV, and the most
useful for determining elemental abundances.  The strength of these
lines depends on the abundance of the elements and their ionization
balance, which in turn depends on the temperature of the cluster.  The
deep gravitational potential well of clusters heats the gas and leaves
it highly ionized.  The gas emits primarily by thermal bremsstrahlung,
and for the temperature range of galaxy clusters the ionization
balance is such that most elements have a large population of their
atoms in the H-like and/or He-like ionization states over most of the
cluster volume.  Clusters are optically thin and nearly isothermal,
with the result that the line emission is easily interpreted without
complicating factors such as radiative transport and the imprint of
non-thermal emission.

While all the elements from carbon to nickel have their main lines in
the X-ray band, not all of them are easily visible.  Elements like
fluorine and sodium have abundances more than an order of magnitude
below the more abundant elements such as silicon and sulfur, and are
so far not detected in observations of galaxy clusters.  The list
below introduces the more abundant and important elements, and the
prospects for measuring their X-ray lines with \asca\ in galaxy
clusters.

\subsection{Carbon, Nitrogen and Oxygen} 

Low temperature clusters and groups may have nitrogen and carbon
K-$\alpha$ lines with significant equivalent width.  However, these
lines lie below the usable bandpass of the \asca\ detectors.

H-like oxygen has strong lines at 0.65\,keV and is an important
element for constraining enrichment scenarios because it is produced
predominantly by type II supernovae.  However, the response of the
\asca\ GIS detector is uncertain at these energies, and the efficiency
of the SIS detector varies with time at low energies and is also
relatively uncertain.  Unfortunately, the usable bandpass we adopt for
\asca\ does not go low enough to include oxygen.

\subsection{Neon and Magnesium}
The K-$\alpha_1$ H-like line for neon is at 1.02\,keV and falls right
in the middle of the iron L-shell complex ranging from about
0.8--1.4\,keV.  The resolution of \asca\ and the close spacing of the
iron lines makes neon abundance determinations from the K-shell
unreliable.

With its K-shell lines also lying in the iron L-shell complex
(1.47\,keV), magnesium suffers from the same problems as neon and is
not well determined with \asca\ data.  Results for both neon and
magnesium from the high resolution RGS on \xmm\ show that the CCD
abundances do not match those obtained with higher resolution gratings
\citep{sak02}, indicating that CCD abundances such as those obtained
from \asca\ are not capable of giving acceptable results.

\subsection{Aluminum}
Aluminum has a higher solar abundance than calcium and argon (the two
lowest abundance elements considered in this paper), but its H-like
K$_\alpha$ line is blended with the much stronger silicon He-like
K$_\alpha$ line and is not reliably measurable.

\subsection{Silicon and Sulfur}
After iron, the silicon abundance is the next most well-determined of
all the elements.  Its H-like K-$\alpha_1$ line at 2.00\,keV and
He-like lines at 1.86\,keV lie in a relatively uncrowded part of the
spectrum, and silicon's large equivalent width leads to a well
determined abundance.

Next to Fe and Si, the high natural abundance of sulfur and its
position in an uncrowded part of the X-ray spectrum make it a well
determined element.  Its K-$\alpha_1$ H-like line is at 2.62\,keV.

\subsection{Argon and Calcium}
The natural abundance of argon is down almost an order of magnitude
from sulfur, giving it a lower equivalent width.  However, the
K-$\alpha_1$ H-like line at 3.32\,keV is in a clear part of the
spectrum and measurable.

Calcium is similar to Ar, with a K-$\alpha_1$ H-like line at
4.10\,keV.

\subsection{Iron and Nickel}
Iron has the strongest set of lines observable in the X-ray spectrum.
High temperature clusters above 3\,keV primarily have as their
strongest lines the K-$\alpha$ set at about 6.97 and 6.67\,keV for
H-like and He-like iron, while lower temperature clusters excite the
L-shell complex of many lines between about 0.6 and 2.0\,keV.
\cite{hwang99} have shown that \asca\ determinations of iron
abundances from just the L or K-shells give consistent results.  Iron
and nickel are predominantly produced by \sno.

Like iron, nickel also has L-shell lines that lie in the X-ray
band. But unlike iron, the abundance determinations are driven almost
entirely by the He-like and H-like K-shell lines at 7.77 and
8.10\,keV. This is because the abundance of nickel is about an order
of magnitude less than iron, and the nickel L-shell lines are blended
with iron's.  Nickel abundances using the H-like and He-like lines are
most reliable for temperatures above $\sim$4\,keV since there is
little excitation of the K-shell line below this energy and because
the reference data for the L-shell lines is not well constrained.

\section{SOLAR ABUNDANCES}

There has been some controversy in the literature as to the canonical
values to use for the solar elemental abundances.  The values for the
elemental abundances by number that are found by spectral fitting to
cluster data do not depend on the chosen values for the solar
abundances.  However, for the sake of convenience elemental abundances
are often reported with respect to the solar values.

\citet{ml96} in their paper report cluster abundances with respect to
the photospheric values in \citet{angr89}.  In \citet{angr89}, the
authors comment on how the photospheric and meteoritic values for the
solar abundances were coming into agreement with better measurement
techniques and improved values of physical constants, and give numbers
for both the photospheric and meteoritic values.  While almost all the
elements were in good agreement, the iron abundance still showed
discrepancies between the photospheric and meteoritic values.
\citet{ia97} questioned the claims in \citet{ml96} by noticing that
they used the photospheric values when analyzing the data (the default
in {\tt XSPEC} then and now), but that the theoretical results they
were comparing to used the meteoritic abundances. Since the
discrepancy in the two values for iron was significant, and because
many of the abundance ratios used in the analysis were with respect to
iron, the conclusions were based on incompatible iron data.

\renewcommand{\arraystretch}{1.0}
\begin{deluxetable}{lcc}
\tablecaption{Solar Abundances\label{solar_abun}}\tablewidth{0pt}
\tablehead{
\colhead{Element} & \colhead{Anders \& Grevesse} &
\colhead{Grevesse \& Sauval}\\
\colhead{} & \colhead{(1989)\tablenotemark{a}} &
\colhead{(1998)\tablenotemark{b}}
}
\startdata
H	& 12.00   	& 12.000 	\\
C	& \phn8.56	& \phn8.520	\\	
N	& \phn8.05	& \phn7.920	\\
O	& \phn8.93	& \phn8.690	\\
Ne	& \phn8.09	& \phn8.080	\\
Mg	& \phn7.58	& \phn7.580	\\
Si	& \phn7.55	& \phn7.555	\\
S	& \phn7.21	& \phn7.265	\\
Ar	& \phn6.56	& \phn6.400	\\
Ca	& \phn6.36	& \phn6.355	\\
Fe	& \phn7.67	& \phn7.500	\\
Ni	& \phn6.25	& \phn6.250
\enddata
\tablerefs{(1) \citealt{angr89}; (2) \citealt{grsa98}.}
\tablecomments{Abundances are given on a logarithmic scale where H is 12.0.}
\tablenotetext{a}{These numbers are the photospheric values, used as
the default in {\tt XSPEC}.}
\tablenotetext{b}{These numbers are a straight average of the
photospheric and meteoritic values (except for oxygen, which has the
updated value given in \citealt{apla01}).}
\end{deluxetable}

Since 1989, the situation has improved.  Reanalysis of the stellar
photospheric data for iron that includes lines from \ion{Fe}{2} in
addition to \ion{Fe}{1} as well as improved modeling of the solar
lines \citep{grsa99} have brought the meteoritic and photospheric
values into agreement. \citet{grsa98} incorporate these changes and
others and has become the \textit{de facto} standard for the standard
solar composition.  Table~\ref{solar_abun}) gives the abundances from
both sources.

However, the past history of changes in the adopted solar composition
implies that there might still be changes in the abundance values for
some elements.  Because of this, and the problems of comparing results
produced with different, incompatible solar values, we quote our
results for the elemental abundances by number with respect to
hydrogen.  We also give the abundances with respect to the
\citet{angr89} solar abundances to ease comparisons with previous
works, and in addition list our results with respect to the standard
\citet{grsa98} values for convenience and for constructing abundance
ratios.

\section{OBSERVATIONS AND DATA REDUCTION}

\subsection{Sample Selection}

We use for our sample all the cluster observations in the archives of
the \asca\ satellite.  In \citet{horner03}\footnote{The results in
\citet{horner03} are primarily from Don Horner's Ph.D. dissertation
\citep{horner01}, found online at: {\tt
http://sol.stsci.edu/\~{}horner}} (hereafter ACC for \asca\ Cluster
Catalog), we describe our efforts to prepare a large catalog of
homogeneously analyzed cluster temperatures, luminosities and overall
metal abundances from the {\tt rev2} processing of the \asca\ cluster
observations.  There we give the full details of the data selection
and reduction; only a brief summary is given here.  In this paper we
use the ACC sample, but our focus is the determination of the
abundances for individual elements in addition to iron.

The \asca\ satellite was launched in February 1993, and ceased
scientific observations in July 2000. Over the course of its lifetime
it observed 434 clusters in 564 observations.  The cluster sample
prepared in ACC selects 273 clusters based on the suitability of the
data for spectral analysis by removing clusters with too few photons
to form an analyzable spectrum, clusters dominated by AGN emission,
etc, and is the largest catalog of cluster temperatures, luminosities,
and abundances. However, because the catalog was designed to maximize
the number of clusters obtained from the \asca\ archives, it is not
necessarily complete to any flux or redshift and could be biased
because of the particular selections of the individual \asca\
observers who originally obtained the data.

The cluster extraction regions in the ACC sample were selected to
contain as much flux as possible in order to best represent the total
emission of the cluster.  Radial profiles of the GIS image were made
and the spectral extraction regions extended out to the point where
the cluster emission was 5$\sigma$ times the background level.
Standard processing was applied to the event files (see ACC).  For the
GIS detector, we used the standard RMF and generated an ARF file for
each cluster. For the SIS detector, we generated RMFs for each chip
and an overall ARF for each cluster.  Backgrounds for the GIS were
taken from the HEASARC blank sky fields except for low galactic
latitude sources $(|b| < 20^\circ)$ where local backgrounds were used.
For the SIS, local backgrounds were used unless the cluster emission
filled the field of view.  In ACC, clusters with more than one
observation had their spectral files combined before analysis; the
joint fitting procedure described below allows us to deal with
multiple observations without a problem.

\subsection{Stacking Analysis}

\begin{figure*}
\begin{center}
\resizebox{\textwidth}{!}{\rotatebox{90}{\includegraphics{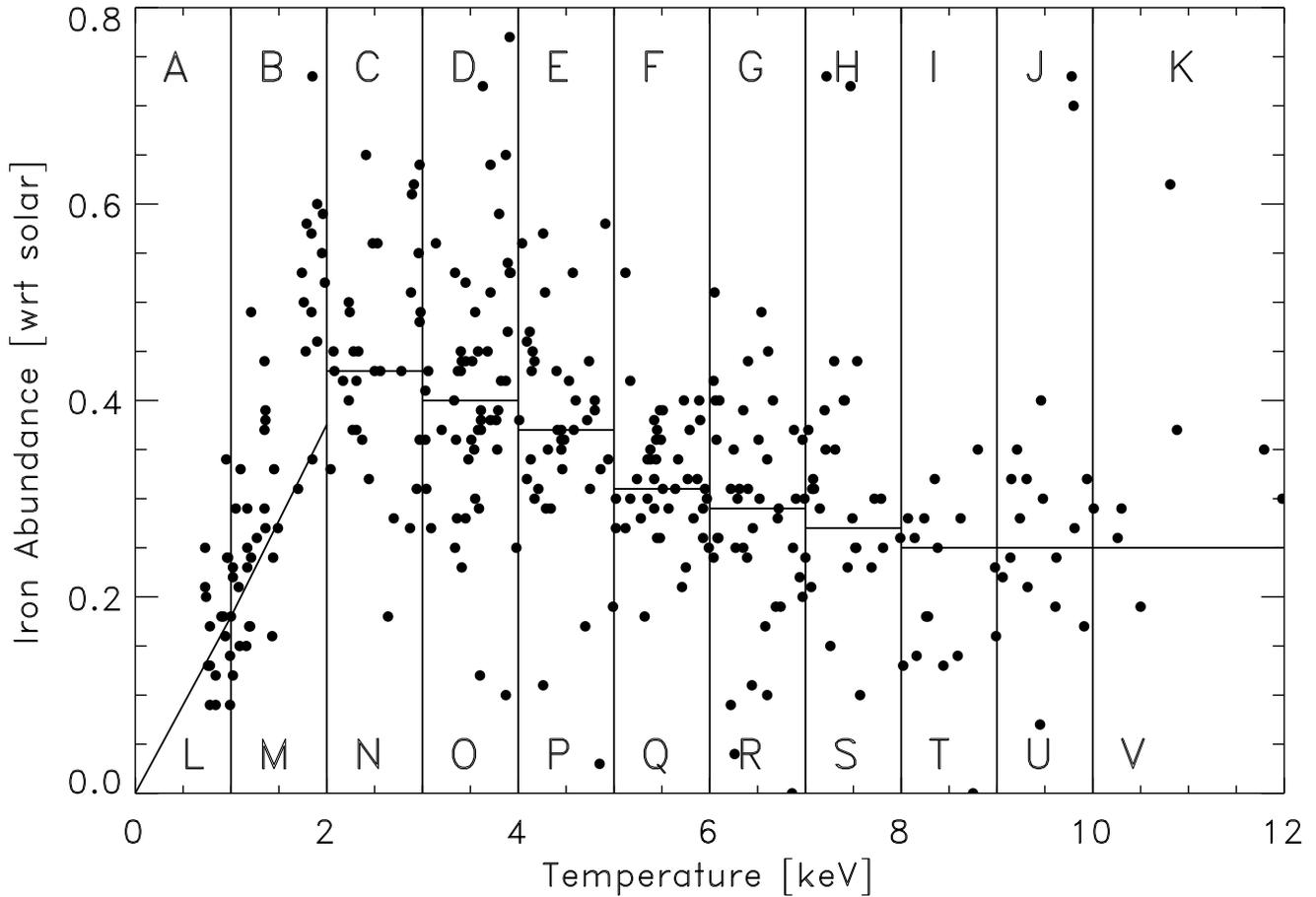}}}
\end{center}
\caption{Stack selection diagram.  The lines on this
abundance--temperature plot show where the boundaries were placed for
the individual stacks. Each point is a single cluster measurement from
ACC.  Stacks A through V were fit individually in {\tt XSPEC}, and the
results from the low and high abundance stacks were combined into a
single result for each one-keV bin (\textit{e.g.} stack A combined with stack L,
etc).
\label{stack_selection}}
\end{figure*}

Only the very few brightest cluster observations in the \asca\
archives have enough signal to noise to allow spectral fitting of the
intermediate elements.  In order to improve our sensitivity to these
elements, we jointly fit a large number of clusters simultaneously.
We divide the 353 observations (of 273 clusters) in the ACC into
fitting groups called stacks by placing together clusters with similar
temperatures and overall abundances (see
Figure~\ref{stack_selection}).  Not only does this allow us to
increase our signal to noise level, but also decreases our
susceptibility to systematic errors resulting from inaccuracies in the
instrument calibration.  Because clusters with different redshifts
(but similar temperatures) are analyzed jointly, any energy-localized
error in the instrument calibration will have less effect on abundance
determination because the error is not likely to affect all the
clusters in a single stack. This method of analysis also smoothes over
biases that may result from the different physical conditions found in
clusters (\textit{e.g.}, substructure and eccentricity).

\begin{figure}[!t]
\resizebox{\columnwidth}{!}{\rotatebox{90}{\includegraphics{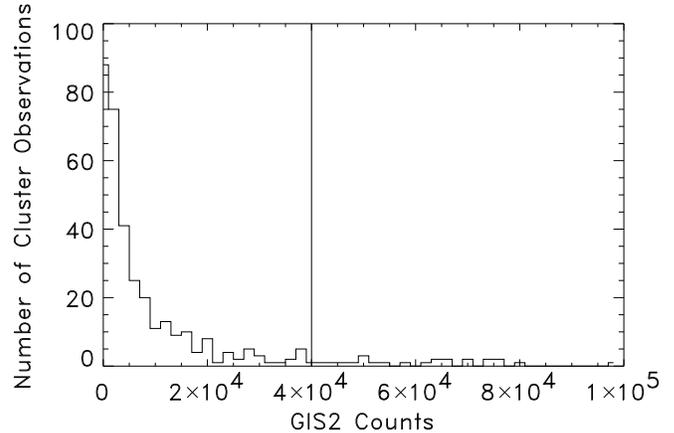}}}
\caption{Counts per cluster histogram.  We excluded clusters with more
than 40\,k counts in the GIS2 detector from our analysis so that very
bright clusters do not unduly bias the joint spectral fitting results.
The vertical line shows the sample cut.  There are only 25 cluster
observations excluded from the analysis because they have too many
counts.
\label{stack_cut}}
\end{figure}

Clusters with more than 40\,k counts in GIS2 were not included in our
analysis so that very bright clusters do not unduly bias the results
for a particular stack.  Figure~\ref{stack_cut} shows a histogram of
the number of counts per cluster observation, and indicates that we
only lose 25 data sets by excluding observations with more than 40\,k
counts.

\begin{deluxetable}{cccr}
\tablecolumns{4}
\tablecaption{Stack Parameters\label{stack_info}}\tablewidth{0pt}
\tablehead{
\colhead{Stack} & \colhead{Temperature} & \colhead{Number of} & \colhead{Total} \\\colhead{Name} & \colhead{Bin (keV)} & \colhead{Clusters} & \colhead{Counts} 
}
\startdata
A   &   0.5 &   17 &   50802\\
B   &   1.5 &   44 &  228685\\
C   &   2.5 &   35 &  261267\\
D   &   3.5 &   47 &  478274\\
E   &   4.5 &   38 &  277014\\
F   &   5.5 &   37 &  391047\\
G   &   6.5 &   39 &  261484\\
H   &   7.5 &   20 &  111593\\
I   &   8.5 &   14 &   93481\\
J   &   9.5 &   13 &  171669\\
K   &  10.5 &   22 &  135321
\enddata
\end{deluxetable}

The number of stacks was motivated by our desire to have a reasonable
number of stacks covering the 1--10\,keV temperature range, and by the
limitations of the {\tt XSPEC} fitting program (jointly fitting more
than about 20 clusters with a variable abundance model exceeds the
number of free parameters allowed).  We divide the clusters into 22
stacks by first separating them into one keV bins (0--1\,keV,
1--2\,keV, \ldots, 9--10\,keV, $10+$\,keV), and then dividing each one
keV bin into a high and low abundance stack using the iron abundances
from ACC.  The split between high and low abundance was made such that
there are roughly an equal number of clusters in the high and low
stack for each one keV bin.  In the case where there are more clusters
in a stack than it is possible to jointly fit, we divide the stack in
two and recombine the results for each sub-stack after fitting.
Table~\ref{stack_info} and Figure~\ref{stack_selection} show the
number of clusters in each stack and the dividing line between high
and low abundance stacks, as well as how the stacks fill the
abundance--temperature plane.

We jointly fit the clusters in each stack to a variable abundance
model modified by galactic absorption ({\tt tbabs*vapec}
\citep{wam,smith01} within {\tt XSPEC}).  The vapec model uses the
line lists of the APEC code to generate a plasma model with variable
abundances for He, C, N, O, Ne, Mg, Al, Si, S, Ar, Ca, Fe, and Ni.  We
fixed He, C, N, O, and Al at their solar values and allowed the other
elements to vary independently (except for Ne and Mg, which we tied
together).  After separately extracting each detector, we combine the
two GIS data sets and the two SIS data sets together before fitting.
The redshift for each cluster was fixed to the optical value found in
the literature, except for the few clusters without published optical
data (which we allowed to vary [see ACC]).  The column density was
fixed at the galactic value for the GIS detectors, but allowed to
float for the SIS detectors in order to compensate for a varying low
energy efficiency problem \citep{yaqoob00}\footnote{ASCA GOF
Calibration Memo (ASCA-CAL-00-06-01, v1.0 06/05/00) \citep{yaqoob00}
can be found at: {\tt
http://heasarc/docs/asca/calibration/nhparam.html}}.  The data for
most clusters was fit between 0.8--10.0\,keV in the GIS and
0.6--10.0\,keV in the SIS.  Observations made after 1998 had a higher
SIS low energy bound of 0.8\,keV because of the low energy efficiency
problem, and about 10 other clusters had modified energy ranges
because of problems with the particular observation (see ACC).

After fitting each of the 22 stacks, we combined the results from the
low and high metallicity stacks in the same temperature range in order
to further improve the statistics.  The difference in the several
elemental abundances between the low and high metallicity stacks was
consistent with an overall higher or lower metallicity (\textit{ie},
the low metallicity stacks L--M have slighter lower Fe, Si, S, and Ni
than the high metallicity stacks A--K).  For example, stack~A (the
0--1\,kev high metallicity stack) was combined with stack~L (the
0--1\,keV low metallicity stack) into a new stack~A.  These final
results for 11 stacks are given in the next section.

\section{RESULTS FOR INDIVIDUAL ELEMENTS}
\label{results}

\renewcommand{\arraystretch}{1.3}
\begin{deluxetable*}{crrrrrrrrr}
\tablecaption{Galaxy Cluster Elemental Abundances by Number\label{abun_table_number}}
\tablewidth{0pt}
\tablehead{
\colhead{Temperature} & \colhead{kT\tablenotemark{a}} & \colhead{Silicon\tablenotemark{b}} & \colhead{Sulfur\tablenotemark{b}} & \colhead{Argon\tablenotemark{b}} & \colhead{Calcium\tablenotemark{b}} & \colhead{Iron\tablenotemark{b}} & \colhead{Nickel\tablenotemark{b}} \\\colhead{Bin (keV)} & \colhead{(keV)}
}
\startdata
  0.5 & $ 0.83_{ 0.82}^{ 0.84}$ & $95.8_{84.4}^{107.9}$ & $107.9_{94.7}^{122.0}$ & $154.8_{134.4}^{177.5}$ & $263.1_{224.0}^{305.0}$ & $102.0_{96.8}^{107.1}$ & $ 0.1_{ 0.0}^{ 1.3}$\\
  1.5 & $ 1.14_{ 1.13}^{ 1.14}$ & $132.3_{125.6}^{138.4}$ & $71.7_{67.1}^{76.1}$ & $39.2_{35.3}^{43.2}$ & $52.8_{47.6}^{58.0}$ & $135.6_{132.4}^{138.9}$ & $ 2.6_{ 0.4}^{ 4.1}$\\
  2.5 & $ 2.58_{ 2.56}^{ 2.60}$ & $163.9_{146.2}^{183.1}$ & $51.6_{43.3}^{60.0}$ & $ 1.0_{ 0.5}^{ 4.7}$ & $ 7.4_{ 4.9}^{10.9}$ & $220.8_{213.8}^{228.7}$ & $ 9.7_{ 7.3}^{13.4}$\\
  3.5 & $ 3.68_{ 3.65}^{ 3.70}$ & $206.1_{188.1}^{225.3}$ & $42.0_{33.1}^{51.2}$ & $ 0.2_{ 0.0}^{ 2.5}$ & $ 1.1_{ 0.0}^{ 3.5}$ & $200.7_{196.4}^{205.8}$ & $20.0_{16.8}^{23.4}$\\
  4.5 & $ 4.57_{ 4.54}^{ 4.61}$ & $197.6_{173.1}^{233.1}$ & $34.2_{20.4}^{50.3}$ & $ 0.0_{ 0.0}^{ 3.7}$ & $ 0.0_{ 0.0}^{ 2.3}$ & $127.7_{121.6}^{131.4}$ & $16.4_{11.9}^{19.7}$\\
  5.5 & $ 5.77_{ 5.72}^{ 5.82}$ & $239.9_{209.0}^{273.6}$ & $16.2_{12.8}^{32.8}$ & $ 0.0_{ 0.0}^{ 2.5}$ & $ 0.0_{ 0.0}^{ 1.8}$ & $137.0_{132.4}^{141.3}$ & $24.5_{20.5}^{28.6}$\\
  6.5 & $ 6.71_{ 6.64}^{ 6.78}$ & $212.5_{182.7}^{254.0}$ & $26.9_{23.4}^{47.7}$ & $ 0.0_{ 0.0}^{ 5.0}$ & $ 0.0_{ 0.0}^{ 1.6}$ & $88.9_{85.1}^{92.6}$ & $16.0_{12.4}^{19.9}$\\
  7.5 & $ 7.45_{ 7.32}^{ 7.60}$ & $158.6_{100.4}^{229.6}$ & $51.2_{21.9}^{84.2}$ & $ 0.0_{ 0.0}^{ 8.4}$ & $ 0.0_{ 0.0}^{ 4.5}$ & $92.1_{85.6}^{99.2}$ & $23.2_{16.8}^{29.8}$\\
  8.5 & $ 8.30_{ 8.12}^{ 8.50}$ & $289.9_{200.8}^{388.2}$ & $66.2_{20.1}^{114.5}$ & $ 2.0_{ 0.0}^{25.6}$ & $ 1.1_{ 0.0}^{10.2}$ & $74.8_{66.4}^{82.8}$ & $21.8_{14.0}^{29.6}$\\
  9.5 & $ 9.63_{ 9.47}^{ 9.80}$ & $364.0_{285.3}^{446.4}$ & $52.2_{24.5}^{92.8}$ & $ 0.3_{ 0.0}^{ 7.5}$ & $ 0.0_{ 0.0}^{ 4.4}$ & $106.2_{98.7}^{113.7}$ & $25.8_{18.7}^{32.9}$\\
 10.5 & $10.92_{10.69}^{11.19}$ & $298.0_{198.7}^{405.9}$ & $64.2_{26.1}^{113.0}$ & $ 0.0_{ 0.0}^{12.4}$ & $ 0.0_{ 0.0}^{ 5.8}$ & $85.1_{76.7}^{93.5}$ & $21.3_{13.6}^{28.9}$
\enddata
\tablenotetext{a}{The values in the temperature column are the fitted temperature of the simultaneously fit clusters in this temperature bin.}
\tablenotetext{b}{All abundances are $1 \times 10^7$ times the number of atoms per hydrogen atom.}
\tablecomments{The numbers in the sub and superscripts for the
abundances are the low and high extent of the 90\% confidence region
for that element.}
\end{deluxetable*}
\renewcommand{\arraystretch}{1.0}

\renewcommand{\arraystretch}{1.3}
\begin{deluxetable*}{cccccccc}
\tablecaption{Classical Galaxy Cluster Elemental Abundances\tablenotemark{a}\label{abun_table_angr}}
\tablewidth{0pt}
\tablehead{
\colhead{Stack} & \colhead{Temperature} & \colhead{Silicon\tablenotemark{a}} & \colhead{Sulfur\tablenotemark{a}} & \colhead{Argon\tablenotemark{a}} & \colhead{Calcium\tablenotemark{a}} & \colhead{Iron\tablenotemark{a}} & \colhead{Nickel\tablenotemark{a}} \\\colhead{Name} & \colhead{Bin} & 
}
\startdata
A   & $  0.5$ & $ 0.27_{ 0.24}^{ 0.30}$ & $ 0.67_{ 0.58}^{ 0.75}$ & $ 4.26_{ 3.70}^{ 4.89}$ & $11.48_{ 9.78}^{13.31}$ & $ 0.22_{ 0.21}^{ 0.23}$ & $ 0.01_{ 0.00}^{ 0.07}$\\
B   & $  1.5$ & $ 0.37_{ 0.35}^{ 0.39}$ & $ 0.44_{ 0.41}^{ 0.47}$ & $ 1.08_{ 0.97}^{ 1.19}$ & $ 2.30_{ 2.08}^{ 2.53}$ & $ 0.29_{ 0.28}^{ 0.30}$ & $ 0.14_{ 0.02}^{ 0.23}$\\
C   & $  2.5$ & $ 0.46_{ 0.41}^{ 0.52}$ & $ 0.32_{ 0.27}^{ 0.37}$ & $ 0.03_{ 0.01}^{ 0.13}$ & $ 0.32_{ 0.21}^{ 0.48}$ & $ 0.47_{ 0.46}^{ 0.49}$ & $ 0.54_{ 0.41}^{ 0.75}$\\
D   & $  3.5$ & $ 0.58_{ 0.53}^{ 0.64}$ & $ 0.26_{ 0.20}^{ 0.32}$ & $ 0.01_{ 0.00}^{ 0.07}$ & $ 0.05_{ 0.00}^{ 0.15}$ & $ 0.43_{ 0.42}^{ 0.44}$ & $ 1.13_{ 0.94}^{ 1.32}$\\
E   & $  4.5$ & $ 0.56_{ 0.49}^{ 0.66}$ & $ 0.21_{ 0.13}^{ 0.31}$ & $ 0.00_{ 0.00}^{ 0.10}$ & $ 0.00_{ 0.00}^{ 0.10}$ & $ 0.27_{ 0.26}^{ 0.28}$ & $ 0.92_{ 0.67}^{ 1.11}$\\
F   & $  5.5$ & $ 0.68_{ 0.59}^{ 0.77}$ & $ 0.10_{ 0.08}^{ 0.20}$ & $ 0.00_{ 0.00}^{ 0.07}$ & $ 0.00_{ 0.00}^{ 0.08}$ & $ 0.29_{ 0.28}^{ 0.30}$ & $ 1.38_{ 1.15}^{ 1.61}$\\
G   & $  6.5$ & $ 0.60_{ 0.52}^{ 0.72}$ & $ 0.17_{ 0.14}^{ 0.29}$ & $ 0.00_{ 0.00}^{ 0.14}$ & $ 0.00_{ 0.00}^{ 0.07}$ & $ 0.19_{ 0.18}^{ 0.20}$ & $ 0.90_{ 0.70}^{ 1.12}$\\
H   & $  7.5$ & $ 0.45_{ 0.28}^{ 0.65}$ & $ 0.32_{ 0.14}^{ 0.52}$ & $ 0.00_{ 0.00}^{ 0.23}$ & $ 0.00_{ 0.00}^{ 0.20}$ & $ 0.20_{ 0.18}^{ 0.21}$ & $ 1.31_{ 0.95}^{ 1.68}$\\
I   & $  8.5$ & $ 0.82_{ 0.57}^{ 1.09}$ & $ 0.41_{ 0.12}^{ 0.71}$ & $ 0.06_{ 0.00}^{ 0.70}$ & $ 0.05_{ 0.00}^{ 0.45}$ & $ 0.16_{ 0.14}^{ 0.18}$ & $ 1.23_{ 0.79}^{ 1.67}$\\
J   & $  9.5$ & $ 1.03_{ 0.80}^{ 1.26}$ & $ 0.32_{ 0.15}^{ 0.57}$ & $ 0.01_{ 0.00}^{ 0.21}$ & $ 0.00_{ 0.00}^{ 0.19}$ & $ 0.23_{ 0.21}^{ 0.24}$ & $ 1.45_{ 1.05}^{ 1.85}$\\
K   & $ 10.5$ & $ 0.84_{ 0.56}^{ 1.14}$ & $ 0.40_{ 0.16}^{ 0.70}$ & $ 0.00_{ 0.00}^{ 0.34}$ & $ 0.00_{ 0.00}^{ 0.25}$ & $ 0.18_{ 0.16}^{ 0.20}$ & $ 1.20_{ 0.76}^{ 1.62}$
\enddata
\tablenotetext{a}{All abundances are with respect to the solar
photosphere elemental abundances given in \citealt{angr89}.}
\tablecomments{The numbers in the sub and superscripts for the
abundances are the low and high extent of the 90\% confidence region
for that element.}
\end{deluxetable*}
\renewcommand{\arraystretch}{1.0}

\renewcommand{\arraystretch}{1.3}
\begin{deluxetable*}{cccccccc}
\tablecaption{Current Galaxy Cluster Elemental Abundances\tablenotemark{a}\label{abun_table_grsa}}
\tablewidth{0pt}
\tablehead{
\colhead{Stack} & \colhead{Temperature} & \colhead{Silicon\tablenotemark{a}} & \colhead{Sulfur\tablenotemark{a}} & \colhead{Argon\tablenotemark{a}} & \colhead{Calcium\tablenotemark{a}} & \colhead{Iron\tablenotemark{a}} & \colhead{Nickel\tablenotemark{a}} \\\colhead{Name} & \colhead{Bin} & 
}
\startdata
A   &   0.5 & $ 0.27_{ 0.24}^{ 0.30}$ & $ 0.59_{ 0.51}^{ 0.66}$ & $ 6.16_{ 5.35}^{ 7.07}$ & $11.62_{ 9.89}^{13.47}$ & $ 0.32_{ 0.31}^{ 0.34}$ & $ 0.01_{ 0.00}^{ 0.07}$\\
B   &   1.5 & $ 0.37_{ 0.35}^{ 0.39}$ & $ 0.39_{ 0.36}^{ 0.41}$ & $ 1.56_{ 1.41}^{ 1.72}$ & $ 2.33_{ 2.10}^{ 2.56}$ & $ 0.43_{ 0.42}^{ 0.44}$ & $ 0.14_{ 0.02}^{ 0.23}$\\
C   &   2.5 & $ 0.46_{ 0.41}^{ 0.51}$ & $ 0.28_{ 0.24}^{ 0.33}$ & $ 0.04_{ 0.02}^{ 0.19}$ & $ 0.32_{ 0.21}^{ 0.48}$ & $ 0.70_{ 0.68}^{ 0.72}$ & $ 0.54_{ 0.41}^{ 0.75}$\\
D   &   3.5 & $ 0.57_{ 0.52}^{ 0.63}$ & $ 0.23_{ 0.18}^{ 0.28}$ & $ 0.01_{ 0.00}^{ 0.10}$ & $ 0.05_{ 0.00}^{ 0.15}$ & $ 0.63_{ 0.62}^{ 0.65}$ & $ 1.13_{ 0.94}^{ 1.32}$\\
E   &   4.5 & $ 0.55_{ 0.48}^{ 0.65}$ & $ 0.19_{ 0.11}^{ 0.27}$ & $ 0.00_{ 0.00}^{ 0.15}$ & $ 0.00_{ 0.00}^{ 0.10}$ & $ 0.40_{ 0.38}^{ 0.42}$ & $ 0.92_{ 0.67}^{ 1.11}$\\
F   &   5.5 & $ 0.67_{ 0.58}^{ 0.76}$ & $ 0.09_{ 0.07}^{ 0.18}$ & $ 0.00_{ 0.00}^{ 0.10}$ & $ 0.00_{ 0.00}^{ 0.08}$ & $ 0.43_{ 0.42}^{ 0.45}$ & $ 1.38_{ 1.15}^{ 1.61}$\\
G   &   6.5 & $ 0.59_{ 0.51}^{ 0.71}$ & $ 0.15_{ 0.13}^{ 0.26}$ & $ 0.00_{ 0.00}^{ 0.20}$ & $ 0.00_{ 0.00}^{ 0.07}$ & $ 0.28_{ 0.27}^{ 0.29}$ & $ 0.90_{ 0.70}^{ 1.12}$\\
H   &   7.5 & $ 0.44_{ 0.28}^{ 0.64}$ & $ 0.28_{ 0.12}^{ 0.46}$ & $ 0.00_{ 0.00}^{ 0.33}$ & $ 0.00_{ 0.00}^{ 0.20}$ & $ 0.29_{ 0.27}^{ 0.31}$ & $ 1.31_{ 0.95}^{ 1.68}$\\
I   &   8.5 & $ 0.81_{ 0.56}^{ 1.08}$ & $ 0.36_{ 0.11}^{ 0.62}$ & $ 0.08_{ 0.00}^{ 1.02}$ & $ 0.05_{ 0.00}^{ 0.45}$ & $ 0.24_{ 0.21}^{ 0.26}$ & $ 1.23_{ 0.79}^{ 1.67}$\\
J   &   9.5 & $ 1.01_{ 0.79}^{ 1.24}$ & $ 0.28_{ 0.13}^{ 0.50}$ & $ 0.01_{ 0.00}^{ 0.30}$ & $ 0.00_{ 0.00}^{ 0.19}$ & $ 0.34_{ 0.31}^{ 0.36}$ & $ 1.45_{ 1.05}^{ 1.85}$\\
K   &  10.5 & $ 0.83_{ 0.55}^{ 1.13}$ & $ 0.35_{ 0.14}^{ 0.61}$ & $ 0.00_{ 0.00}^{ 0.49}$ & $ 0.00_{ 0.00}^{ 0.25}$ & $ 0.27_{ 0.24}^{ 0.30}$ & $ 1.20_{ 0.76}^{ 1.62}$
\enddata
\tablenotetext{a}{All abundances are with respect to
the average of the photospheric and meteoritic solar elemental
abundances given in Table~\ref{solar_abun} adapted from \citealt{grsa98}.}
\tablecomments{The numbers in the sub and superscripts for the
abundances are the low and high extent of the 90\% confidence region
for that element.}
\end{deluxetable*}
\renewcommand{\arraystretch}{1.0}

\renewcommand{\arraystretch}{1.3}
\begin{deluxetable*}{ccrrrrr}
\tablecaption{Elemental Abundance Ratios\label{abun_table_ratios}}
\tablewidth{0pt}
\tablehead{
\colhead{Stack} & 
\colhead{Temperature} & 
\colhead{[Si/Fe]} & 
\colhead{[S/Fe]} & 
\colhead{[Si/S]} & 
\colhead{[Ni/Fe]} & 
\colhead{[Fe/H]}\\
\colhead{Name} & 
\colhead{Bin}
}
\startdata
A   &   0.5 & $-0.08_{-0.14}^{-0.03}$ & $ 0.26_{ 0.20}^{ 0.32}$ & $-0.34_{-0.42}^{-0.27}$ & $-1.81_{-\infty}^{-0.64}$ & $-0.49_{-0.51}^{-0.47}$\\
B   &   1.5 & $-0.07_{-0.09}^{-0.04}$ & $-0.04_{-0.07}^{-0.01}$ & $-0.02_{-0.06}^{ 0.01}$ & $-0.47_{-1.27}^{-0.26}$ & $-0.37_{-0.38}^{-0.36}$\\
C   &   2.5 & $-0.18_{-0.24}^{-0.13}$ & $-0.40_{-0.47}^{-0.33}$ & $ 0.21_{ 0.12}^{ 0.29}$ & $-0.11_{-0.23}^{ 0.03}$ & $-0.16_{-0.17}^{-0.14}$\\
D   &   3.5 & $-0.04_{-0.08}^{-0.00}$ & $-0.44_{-0.55}^{-0.36}$ & $ 0.40_{ 0.29}^{ 0.49}$ & $ 0.25_{ 0.17}^{ 0.32}$ & $-0.20_{-0.21}^{-0.19}$\\
E   &   4.5 & $ 0.13_{ 0.07}^{ 0.21}$ & $-0.34_{-0.56}^{-0.17}$ & $ 0.47_{ 0.23}^{ 0.65}$ & $ 0.36_{ 0.22}^{ 0.44}$ & $-0.39_{-0.42}^{-0.38}$\\
F   &   5.5 & $ 0.19_{ 0.13}^{ 0.25}$ & $-0.69_{-0.80}^{-0.39}$ & $ 0.88_{ 0.76}^{ 1.19}$ & $ 0.50_{ 0.42}^{ 0.57}$ & $-0.36_{-0.38}^{-0.35}$\\
G   &   6.5 & $ 0.32_{ 0.25}^{ 0.40}$ & $-0.28_{-0.35}^{-0.04}$ & $ 0.61_{ 0.51}^{ 0.86}$ & $ 0.50_{ 0.39}^{ 0.60}$ & $-0.55_{-0.57}^{-0.53}$\\
H   &   7.5 & $ 0.18_{-0.02}^{ 0.34}$ & $-0.02_{-0.39}^{ 0.20}$ & $ 0.20_{-0.29}^{ 0.45}$ & $ 0.65_{ 0.51}^{ 0.76}$ & $-0.54_{-0.57}^{-0.50}$\\
I   &   8.5 & $ 0.53_{ 0.36}^{ 0.67}$ & $ 0.18_{-0.35}^{ 0.42}$ & $ 0.35_{-0.27}^{ 0.61}$ & $ 0.71_{ 0.51}^{ 0.85}$ & $-0.63_{-0.68}^{-0.58}$\\
J   &   9.5 & $ 0.48_{ 0.37}^{ 0.57}$ & $-0.07_{-0.41}^{ 0.18}$ & $ 0.55_{ 0.18}^{ 0.81}$ & $ 0.64_{ 0.49}^{ 0.74}$ & $-0.47_{-0.51}^{-0.44}$\\
K   &  10.5 & $ 0.49_{ 0.30}^{ 0.63}$ & $ 0.11_{-0.29}^{ 0.36}$ & $ 0.38_{-0.12}^{ 0.64}$ & $ 0.65_{ 0.44}^{ 0.78}$ & $-0.57_{-0.62}^{-0.53}$
\enddata
\tablecomments{All abundance ratios are with respect to
the current abundances given in Table~\ref{abun_table_grsa}.  The
numbers in the sub and superscripts for the abundances are the low and
high extent of the 90\% confidence region for that element.
Abundances are given in the usual dex notation, ie: [A/B]$\,\equiv
\rm{log}_{10}
(N_A/N_B)_{cluster}-\rm{log}_{10}(N_A/N_B)_\odot$.}
\end{deluxetable*}
\renewcommand{\arraystretch}{1.0}

We present results for the abundance of the elements Fe, Si, S, Ar,
Ca, and Ni as a function of cluster temperature.  Other elements with
lines present in cluster X-ray spectra (\textit{e.g.} Ne, Mg and O)
have statistical or systematic errors too large to allow meaningful
results.  The main results of our analysis are presented in
Table~\ref{abun_table_number} which lists the metal abundances of the
cluster stacks.  These results are given by number with respect to
hydrogen.  In Table~\ref{abun_table_angr} we give the same results
with respect to the photospheric solar abundances in \cite{angr89} in
order to allow easy comparison with other results in the literature.
Finally, in Table~\ref{abun_table_grsa} we give the cluster abundances
with respect to the standard solar composition given in our
Table~\ref{solar_abun} adopted from \cite{grsa98}.

\begin{figure*}
\begin{center}
\resizebox{!}{8in}{\rotatebox{0}{\includegraphics{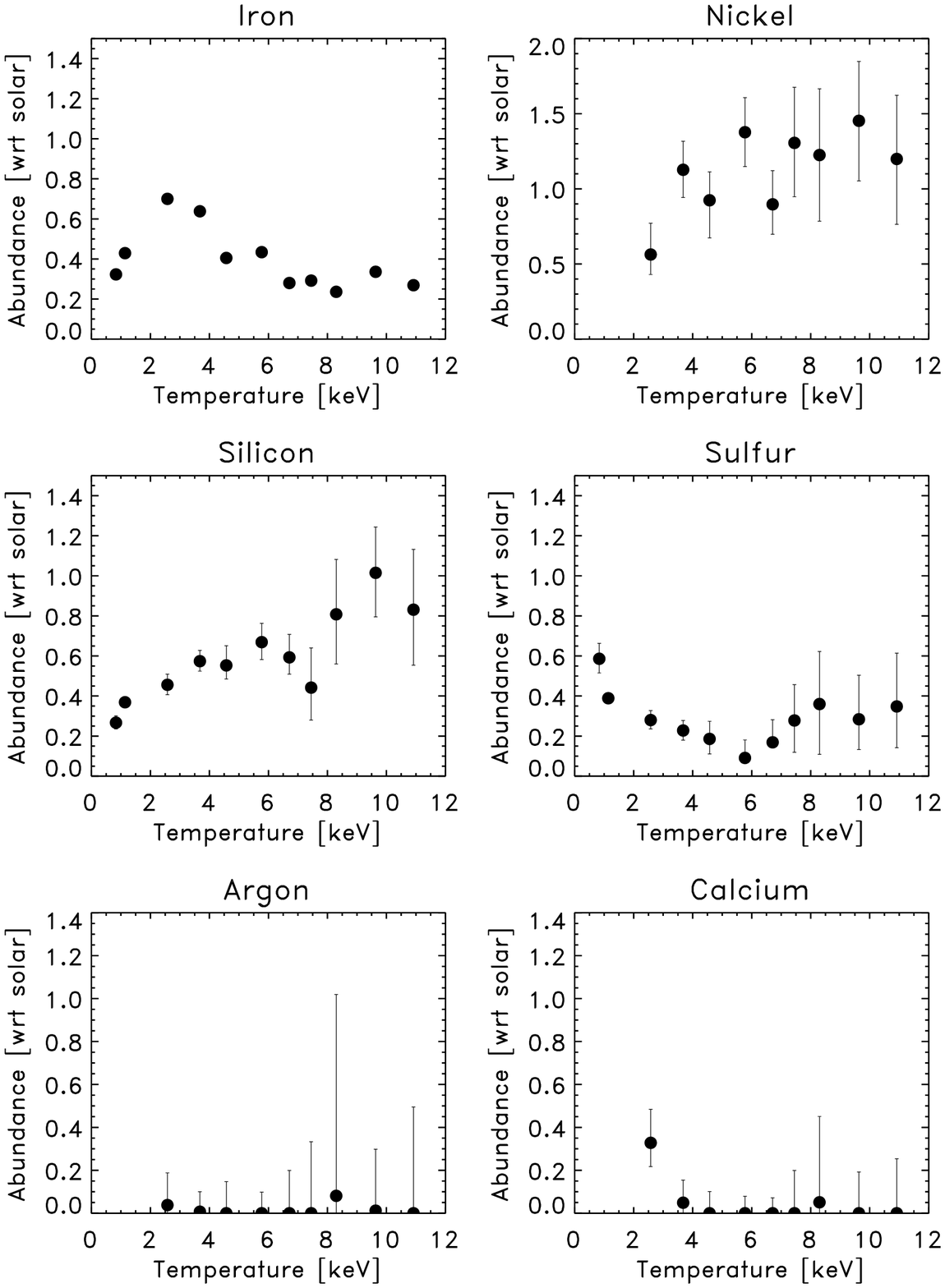}}}
\end{center}
\caption{The galaxy cluster elemental abundances as a function of
temperature.  These abundances are with respect to the solar abundances
of \citet{grsa98}.  The error bars are the 90\% confidence interval
for that elemental abundance; the error bars for iron are smaller than
the plotted points.\label{abuns_fig}}
\end{figure*}
Figure~\ref{abuns_fig} displays the same information as in
Table~\ref{abun_table_grsa}, but in graphical form.  Several
interesting points are immediately apparent in the data.  First, the
results for iron closely follow previous results for cluster
metallicities, but have much smaller error bars.  The actual numerical
result for the iron abundance is consistent with the previous results
of about 1/3 solar, but the improved quality of the data allows the
detection of trends in the iron abundance with temperature: at high
temperatures above 6\,keV, the abundance is constant at a value of 0.3
solar.  Between\,2.5\,keV and 6\,keV, the iron abundance falls from
0.7 solar to 0.3 solar, and from 0.5\,keV to 2.5\,keV the abundance
rises steeply from 0.3 solar to 0.7 solar.

\begin{figure}[!t]
\begin{center}
\resizebox{\columnwidth}{!}{\rotatebox{90}{\includegraphics{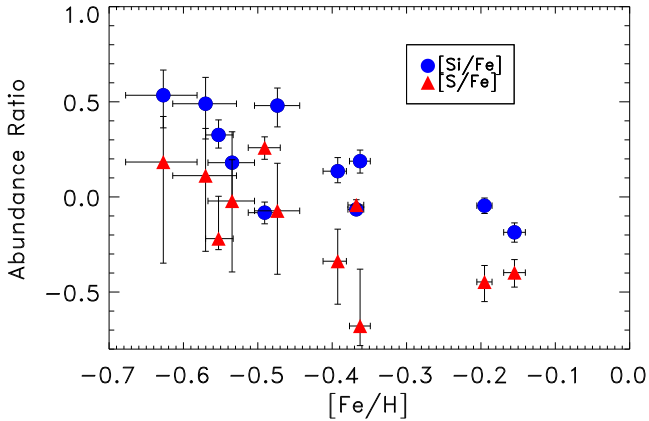}}}
\end{center}
\caption{The silicon and sulfur abundances ratios with respect to
iron.\label{ratios_v_fe_fig}}
\end{figure}

Also of note are the results from the second and third most strongly
detected elements, silicon and sulfur.  Here, the abundance ratios
give some important information: the value of [Si/Fe] is generally
super-solar, while the value of [S/Fe] is sub-solar.
Table~\ref{abun_table_ratios} gives the ratios of silicon, sulfur, and
nickel to iron.  Figure~\ref{ratios_v_fe_fig} shows the trend of the
abundance ratios [Si/Fe] and [S/Fe] as a function of the iron
abundance.
Silicon and sulfur also show disagreement in their trend with
temperature --- the relative abundance of silicon rises with
temperature, and sulfur falls.  We find that the silicon abundance
rises from $\sim$0.3~solar in cooler clusters to $\sim$0.7~solar in
hot clusters, in excellent agreement with the results of
\citet{fuk98}.

Calcium and argon are noticeable for their lack of a detectable signal.
Both are not detected in the stacked \asca\ data, yielding only upper
limits over the temperature range from 2--12\,keV.  

The relative abundances for nickel are measured to be higher than any
other element in our study.  For clusters hotter than 4\,keV, nickel
is about 1.2 times the solar value.  Lower temperature clusters are
too cool to significantly excite the nickel K-$\alpha$ transition, and
abundances derived from the Ni L-shell lines are not reliable. The
Ni/Fe values are extremely high at about 3 times the solar value,
confirming the results of \cite{dwa,da,dwb}.

\section{SYSTEMATIC ERRORS}

The abundances results in this paper depend upon the measurement of
particular spectral lines that are not always very strong.  Because of
this, a proper understanding of the results must take into account any
systematic errors that may bias the results.  Of particular concern to
this work are systematic errors in the calibration of the effective
area, since these often manifest themselves as lines in residual
spectra of sources with a smooth continuum.  If these residual lines
fall at the same energy as the important elemental spectral lines they
can have a significant affect on the derived elemental abundances.

The ACC catalog paper discusses general \asca\ systematics and our
sensitivity to them.  In addition, there are several other tests we
have undertaken to determine the effect of small line-like systematic
errors in the effective area.  In order to determine the size of any
calibration errors, we have fit spectra from broad band continuum
sources and quantified the residuals.  We have also used these
residuals to correct the cluster spectra, and have compared the
derived elemental abundances to those from the uncorrected spectra.

We have used an \asca\ observation of Cygnus~X-1 (a bright source
where the systematic errors dominate the statistical ones) to measure
the size of residual lines in fitted spectra of a continuum source.
These lines could be due to errors in the instrument calibration and
could affect the abundance determinations in clusters.  Using a {\tt
power law + diskline} model in {\tt XPSEC}, we measure the largest
residual in the Cyg~X-1 spectrum to have an equivalent width of 17\,eV
at 3.6\,eV.  Using the Raymond-Smith plasma code to model the
equivalent width of elemental X-ray lines, we find that the only
element with a small enough equivalent width to be possibly affected
by a 17\,eV residual is calcium, with an equivalent width of 25\,eV
for very hot clusters ($>6\,$keV). However, the 17\,eV residual lies
at the wrong energy to affect calcium (lines at 3.8, 3.9, and
4.1\,keV).  In addition, the positive residual would serve to increase
the measured calcium abundances; our measured calcium abundances are
lower than expected.  A similar test with the bright continuum source
Mkn 421 (with the core emission removed to prevent pileup) shows a
maximum residual of 18\,eV equivalent width at the 2.11\,keV Au edge
of the mirror.  These results indicate that any line-like calibration
errors are not large enough to affect the cluster elemental abundance
measurements in this study.

\begin{figure}[!t]
\begin{center}
\resizebox{\columnwidth}{!}{\rotatebox{-90}{\includegraphics{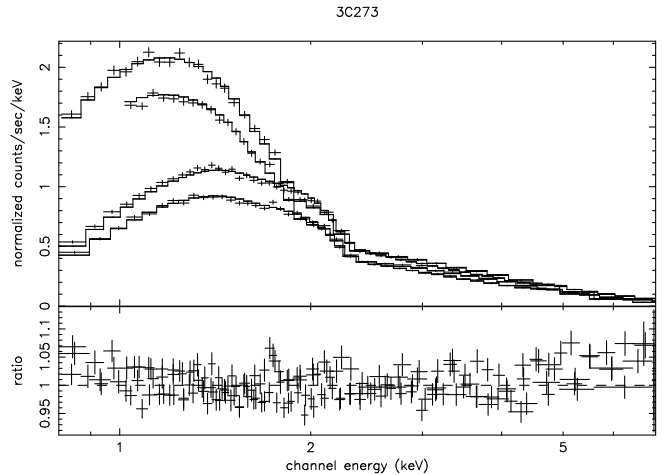}}}
\end{center}
\caption{An absorbed power law fit to the \asca\ 3C\,273 data.  The data
and model are shown in the top panel for the 4 \asca\ detectors, and
the ratio of the data to the model is shown in the bottom panel.  The
lack of any significant residuals indicate that the \asca\ effective
area calibration is free of any significant line-like systematic
errors.
\label{3c273}}
\end{figure}

We have also used an \asca\ observation of 3C\,273 (sequence number
12601000) as a broad band continuum source to check for calibration
errors in the response matrices and their effect on the derived
cluster elemental abundances.  We fit 3C\,273 with an absorbed power
law model (Figure~\ref{3c273})
and extracted the ratio residuals to the best fit.  We then used these
residuals as a correction to the cluster data, and then refit the
clusters to find the abundances.  The abundances from the corrected
data are completely consistent with the abundances from the
uncorrected data, further indicating that errors in the effective area
calibration do not affect the derived elemental abundances.

\begin{figure*}[!t]
\begin{center}
\resizebox{!}{8in}{\rotatebox{0}{\includegraphics{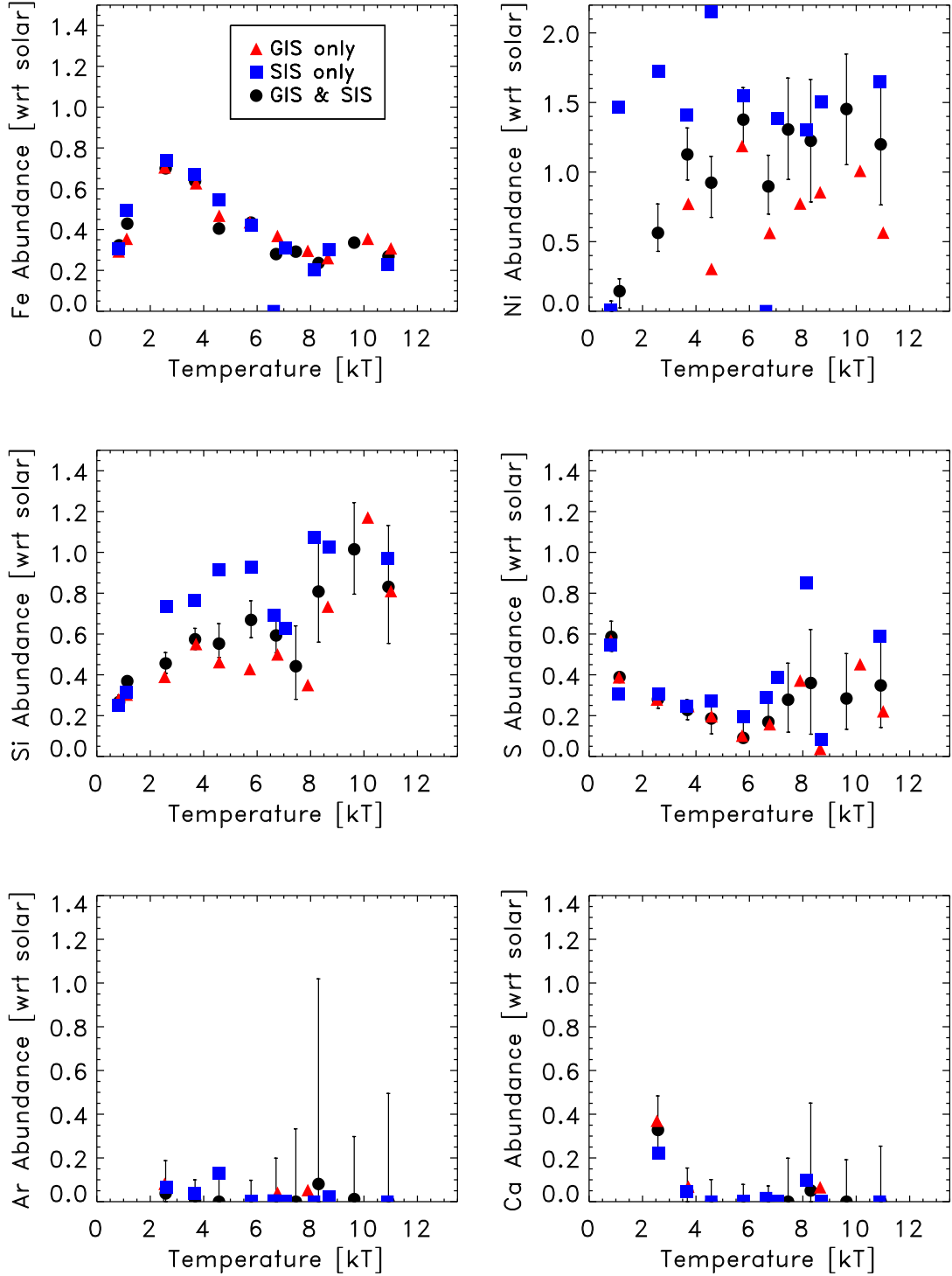}}}
\end{center}
\caption{A comparison of abundance results using the GIS alone, the
SIS alone, and the GIS and SIS combined.  The GIS and SIS agree well
except for nickel and a slight trend in the medium temperature silicon
results.  The nickel discrepancies are due to the lower effectiveness
of the SIS at Ni K, and the SIS silicon measurements suffer from
proximity to the detector Si edge.  The errorbars for the GIS and SIS
alone points are omitted for clarity, but are about a factor of
$\sqrt{2}$ larger than those for the combined points.
\label{gissis}}
\end{figure*}

We have also investigated the abundances derived using the GIS and SIS
detectors separately in order to check the consistency of our results.
While the SIS has better spectral resolution, it also suffers from
slight CTI and low energy absorption problems that do not affect the
GIS.  Figure~\ref{gissis}
shows these results.  In general, the GIS and SIS abundances match
very well for most of the elements.  However, nickel and silicon show
a systematic trend of slightly higher SIS abundances.  The SIS nickel
abundances are not as reliable as the GIS abundances because the GIS
has more effective area than the SIS at Ni K.

The systematic trend in the medium temperature silicon abundances is
more difficult to understand. \cite{fuk97} showed that the GIS and SIS
sulfur and silicon abundances for his cluster sample were well
matched.  Individual analysis of the bright, medium temperature
clusters Abell~496 and Abell~2199 show that the results are indeed
real and verify that the SIS gives higher Si abundances than the GIS.
Additional fits to the data that allow the SIS gain to vary also do
not significantly affect the abundance results.  One possible
contribution to this problem is that the medium temperature clusters
most affected by this trend mostly lie at a redshift such that the Si
K-shell lines (1.86 and 2.00\,keV) are redshifted very close to the Si
edge (1.84\,keV) in the detector.  These silicon abundances are less
reliable because of structure present in the Si edge not well modeled
in the instrument response \citep{mori}.

The fit results in the 0.5 and 1.5\,keV bins from
Tables~\ref{abun_table_number}, \ref{abun_table_angr}, and
\ref{abun_table_grsa} show elevated calcium and argon abundances.
These very high fit results (1--10 times solar) are not believed to be
indicative of the actual cluster abundances because of systematic
errors that preferentially affect these low temperature bins.  Two
sources of systematic error we have identified both become important
at low cluster temperatures and affect the modeled spectrum at the
higher energies of the Ar and Ca K-shell lines.

The first is background subtraction: because the dominant
bremsstrahlung emission is limited to lower energies for the cooler
clusters, the cluster flux present at the higher energies of the Ar
and Ca K-shell lines becomes comparable to the background.  Small
errors in the blank sky background can then significantly affect the
background-subtracted data, leading to incorrect abundances at higher
energies where the relative errors from the background subtraction
become large.  

Also important at the lower cluster temperatures where the
bremsstrahlung emission does not dominate throughout the X-ray
spectrum is the contribution from point sources within the individual
galaxies.  \citet{angelini01} and \citet{irwin03} have shown with data
from elliptical galaxies that the contribution from X-ray binaries is
important and can be characterized with a 7\,keV bremsstrahlung model.
Since we do not include this component in our cluster fits, we expect
that our Ar and Ca abundance for low temperature clusters will be
driven upwards in an attempt to try and match the flux actually
contributed by the X-ray point sources.

\section{COMPARISONS}

We believe that X-ray determinations of elemental abundances in galaxy
clusters are one of the most accessible means for obtaining useful
abundance measurements.  While our results are consistent with and
significantly expand previous cluster X-ray measurements, they do not
always agree with measurements of elemental abundances in other
objects.

\subsection{Comparison to Other Cluster X-ray Measurements}

\subsubsection{\asca\ Measurements}
Measurements of abundances for elements other than iron has
historically been difficult because of the low equivalent width of the
lines and the limited resolution of X-ray detectors.  Aside from a few
limited results with crystal spectrometers on \textit{Einstein} and
other satellites, the first chemical abundance measurements of
elements other than iron in galaxy clusters was made by \citet{ml96}
and made use of the high resolution of the CCD cameras onboard \asca.
Their results for 4 moderate temperature (2.5--5.0\,keV), very bright
clusters show an average Fe abundance of 0.32~solar, 0.65~solar for
Si, 0.25 for S, and 1.0 for Ni.  All of these results are in agreement
with our results for moderate temperature clusters presented in
Table~\ref{abun_table_angr}.  \citet{fuk98} also reported results for
the silicon abundance in clusters that used \asca\ data.  Their
results for 40 clusters was in general agreement with the data from
\citet{ml96}, showing a generally constant iron abundance for clusters
above 3\.keV, and also hinted at the silicon trend with temperature
presented with more detail here.

\citet{f00} and \citet{f01} concentrated on the detection of abundance
gradients with \asca.  Their data for Fe and Si at large radii are in
agreement with ours.  However, they present measurements of S coupled
with Ar in order to reduce errors in the S determination.  We see that
these elements have very different abundances, and can't compare our
data directly to theirs.

\subsubsection{\xmm\ Measurements}
Further results for the intermediate element abundances in clusters
with \asca\ were hampered by its moderate resolution, and especially
because only a handful of clusters have high enough X-ray flux to
allow meaningful measurements.  The advent of \textit{Chandra} and
especially \xmm\ have changed this with their higher resolution CCD
cameras and much increased effective area.  While measurements of
argon and calcium are still beset with systematic problems in the
\xmm\ response and background subtraction, measurements of silicon and
sulfur are largely free of these complications.

The much improved spatial and spectral resolution of \xmm\ allow not
only abundance measurements averaged over the whole cluster, but also
spatially resolved ones.  While our measurements in clusters are
overall spatial averages, the measurements with \xmm\ are usually
spatially resolved and give information on abundance gradients within
clusters. While this information is important and sheds valuable light
on the enrichment mechanisms in clusters, we will limit our
comparisons to overall abundances measured with \xmm, or with
abundances in the more voluminous outer regions of clusters with
spatially resolved abundances.

The many \xmm\ papers on M87 \citep{bohr01,mg01,gm02,mat03} all find a
significant abundance gradient in the center of the cluster.  In the
very center, \citet{mg01} find generally higher abundances than in the
outer regions, with the iron abundance at about 0.6 the solar value.
They find silicon to be 1.0 in the same region, giving [Si/Fe] only
slightly lower than our results for a similar temperature cluster.
However, sulfur is about 1.1 solar, overly abundant with respect to
iron than in our results.  The discrepancies between some of our
results for 2.5\,keV clusters and the \xmm\ M87 results are tempered
by the fact that earlier \asca\ results for M87 \citep{matsu96,gm99}
are similar to the \xmm\ results, and by the fact that the M87 data is
for only one cluster, while our \asca\ results are for an average of
several clusters at the same temperature.

\citet{tamura01} find sulfur and silicon to have similar abundances in
the center of A496 with RGS data from \xmm, slightly different than
our results that show silicon slightly higher than sulfur by a factor
of about 1.5.  Unfortunately, the M87 data concentrates on the cluster
core, and the A496 data is taken with the RGS, which also is limited
to the centers of bright clusters.  This bias towards the very centers
of clusters makes a comparison with our average cluster abundances
difficult.

\citet{f02} also look at \xmm\ data for M87, but focus their paper on
a discussion of the supernovae enrichment needed to account for the
observed abundances.  Their data leads them to conclude, as we do,
that the standard yields of the canonical \sno\ and \snt\ models are
not sufficient to explain the pattern of abundances observed.  Their
solution calls for an additional source of metals from a new class of
type I supernovae.  This scenario does not fit our data well;
specifically, it leads to an overproduction of sulfur, argon and
calcium, with not enough silicon produced to match our observations.
\citet{f02} focus on the inner the inner 70\,kpc region of M87 because
of the cluster's proximity.  Their results in this region are where
the influence of the galaxy's stellar population and associated SNIa
are most keenly apparent. Even at the outer parts of this region, the
abundances show signs of a transition to a more typical cluster
pattern. This inner region contributes a small fraction of the
emission measure of most systems in our ACC sample, and makes
comparisons difficult since we sample different areas of the cluster.

\subsection{Comparison to Measurements in Different Types of Objects}

\subsubsection{The Thin Disk}

\begin{figure}[!t]
\begin{center}
\resizebox{\columnwidth}{!}{\rotatebox{90}{\includegraphics{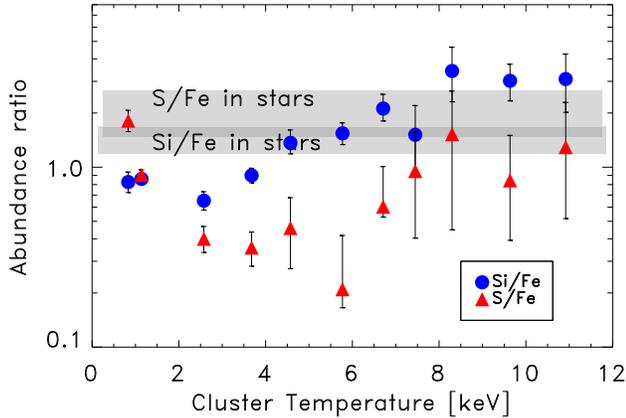}}}
\end{center}
\caption{The silicon and sulfur abundances as a function of
temperature, compared to their abundance in stars.  The upper grey bar
is the S/Fe data from \citet{tww95}, and the overlapping lower grey
bar is the Si/Fe data from the same source.\label{s_si_stars_fig}}
\end{figure}

The standard solar elemental composition is based on measurements
taken of the sun, which resides in the thin disk of the galaxy.  Our
cluster results show significant differences with stellar data from
the thin disk.  Figure~\ref{s_si_stars_fig}
shows silicon and sulfur abundance ratios in clusters compared with
data from \citet{tww95}.  The stellar data is overplotted on the
cluster points as grey bars. The [Si/Fe] cluster data overlaps with
the stellar data, but has a much greater range.  The [S/Fe] cluster
data lies almost totally below the stellar data.  The largest
discrepancy is between the cluster and stellar nickel abundance
ratios; clusters are higher than the stellar data of 0.1 by 0.5 dex.
This very high value for [Ni/Fe] is not seen in stars at any
metallicity.

\subsubsection{The Thick Disk}

The majority of the stellar mass in clusters resides in elliptical
galaxies and the bulges and halos of spirals \citep{bell03}.
\citet{prochaska00} find that the abundance patterns of the thick disk
are in excellent agreement with those in the bulge, suggesting that
the two formed from the same reservoir of gas.  If this is true, and
if the majority of the enriched gas in the ICM originated in these
numerically dominant stellar populations, then we would expect that
the abundance pattern in the ICM is in good agreement with that in the
thick disk of the Milky Way.  However, reality is more
complicated. While \citet{prochaska00} find that the thick disk has
enhanced abundances for the $\alpha$ elements compared to the thin
disk and solar neighborhood, they also find [Ni/Fe] in the thick disk
is similar to its solar value.  Additionally, the calcium abundance
ratio is found to be super solar ([Ca/Fe] = 0.2), and sulfur is more
enhanced with respect to iron in the thick disk than in clusters.
Also, the recent work of \citet{pompeia03} shows a [Si/Fe] value of
nearly 0.0, much lower than what we observe here.

\subsubsection{\ion{H}{2} Regions and Planetary Nebulae}

Comparison of our cluster results with \ion{H}{2} regions or planetary
nebulae is difficult because of the strong effects of dust and non-LTE
conditions on abundance measurements of elements like iron and silicon
\citep{peimbert01,stas02}.  C, N, and O are more readily observed in these
objects, but are not observed with \asca.

\subsubsection{Damped Ly-$\alpha$ Absorbers}

\citet{pw02} show abundance measurements from a large database of
damped Ly-$\alpha$ (DLA) observations.  Their results are for the
redshift range $1.5 < z < 3.5$ and indicate that there is already
significant $\alpha$ element enhancement at high redshifts.  This
conclusion is consistent with that reached by \citet{ml96} in an
analysis of high redshift clusters.  The [Fe/H] value observed by
\citeauthor{pw02} of -1.6 is constant over a wide range of redshift,
indicating that the enrichment was at earlier times.  This value is
much lower than our value of [Fe/H] = -0.5 (for our higher temperature
bins where the iron abundance is fairly constant), showing the
importance of supernovae enrichment in the ICM.  Their average [Si/Fe]
value of 0.35 agrees with our results for moderate temperature
clusters, and their spread of [Si/Fe] measurements from 0.0 to 0.5 is
well matched with ours.  Their [Si/Fe] values are also constant over a
wide redshift range, further supporting early enrichment.  However,
their results for sulfur differ from ours in that they show more
enrichment with respect to iron ([S/Fe] $\sim$ 0.4) than we do.  This
suggests that later enrichment by supernovae into the ICM had a
reduced role for sulfur in comparison to the other elements.  Nickel
is also different in these systems than in clusters, with [Ni/Fe] of
only 0.07; much different than our very high value of $\sim$0.5.
While \citet{pw02} have some results for argon, they are widely
scattered and not easily interpreted.  The differences with the
cluster measurements of S and Ni indicate that the stellar population
that enriched these systems is probably not the origin of the metals
in the ICM.

\subsubsection{Lyman Break Galaxies}

\citet{pettini02} present detailed observations of a lensed Lyman
break galaxy at $z=2.7$.  Their results also indicate that significant
metal enrichment have already taken place at high redshift.  Their
results for O, Mg, Si, S, and P are all at about 0.4 the solar value,
showing the results of fast supernovae processing.  However, the iron
peak elements are not as enriched, with the values of Mn, Fe, and Ni
at only 1/3 solar.  The interpretation is that this galaxy has been
caught in the middle of turning its gas into stars, and that while
enrichment from \snt\ has occurred, not enough time has passed to
allow significant enrichment from the iron peak producing \sno. The
overall metallicity of [Fe/H] = -1.2 is closer to our observed cluster
value than the measurements from the DLAs are.  However, the abundance
ratios in this object are also sufficiently different than in clusters
to indicate that the stellar population that enriched the Lyman break
galaxies is also probably not the source of the metals in the ICM.

\subsubsection{The Ly-$\alpha$ Forest}

While it is possible to probe the metallicity history of the universe
as a function of redshift \citep{song01,song02}, the difficulty
inherent in measuring lines from many different elements in the
Ly-$\alpha$ forest has made abundance measurements of most elements
besides Fe, C, N, and Si unavailable. Ly-$\alpha$ forest measurements
show metallicities of [Fe/H] ranging from -2.65 at a redshift of 5.3 to
-1.25 at redshift 3.8.  All of these are almost two orders of
magnitude lower than our measurements.

\section{SUPERNOVAE TYPE DECOMPOSITION}

With the results of \S~\ref{results} for the abundance of the
intermediate elements in clusters, it is possible to try to constrain
the mix of supernovae types that have enriched the ICM.  In
\S~\ref{2types}, we check to see how well the yields from the
standard \sno\ and \snt\ models can reproduce the cluster
observations.  We find that other sources of metals are necessary.
Then we comment on the expected cluster abundances derived using some
of the standard models in the literature.  In \S~\ref{3types}, we
investigate different SN models that produce intermediate elements
with the abundance ratios necessary for reconciling the models with
the observations.

\subsection{SN Fraction Analysis using Canonical \sno\ and \snt\ Models}
\label{2types}

\begin{figure}[!t]
\begin{center}
\resizebox{\columnwidth}{!}{\rotatebox{90}{\includegraphics{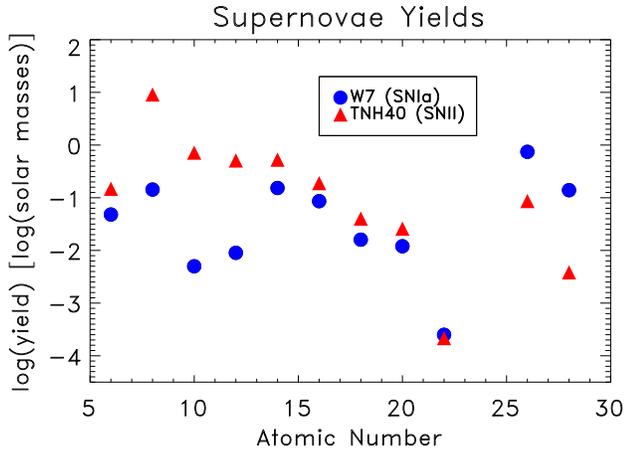}}}
\end{center}
\caption{The Type~Ia and Type~II supernovae yields.  The horizontal
axis is atomic number --- O is 8, Fe 26, Si 14, S 16, Ni 28.  The
yields are in solar masses per supernova.  Iron comes mostly from
\sno\, and nickel even more so. Si, S, Ar, and Ca are a mix of \snt\ and
\sno\, but with a majority \snt\ contribution. The \sno\ data comes
from the W7 model, and the \snt\ data is from the IMF mass-averaged
TNH-40 model.
\label{yields}}
\end{figure}

We use the yields of the revised W7 model \citep{n97} for \sno\ and
the TNH-40 yields \citep{tnh40} (mass-averaged by integrating across
the IMF to an upper limit of 40 solar masses) for \snt\ (shown in
Figure~\ref{yields})
as our basis.  Each model includes the amount of each of the
intermediate elements produced in a supernova of that type.

\begin{figure}[!t]
\begin{center}
\resizebox{\columnwidth}{!}{\rotatebox{90}{\includegraphics{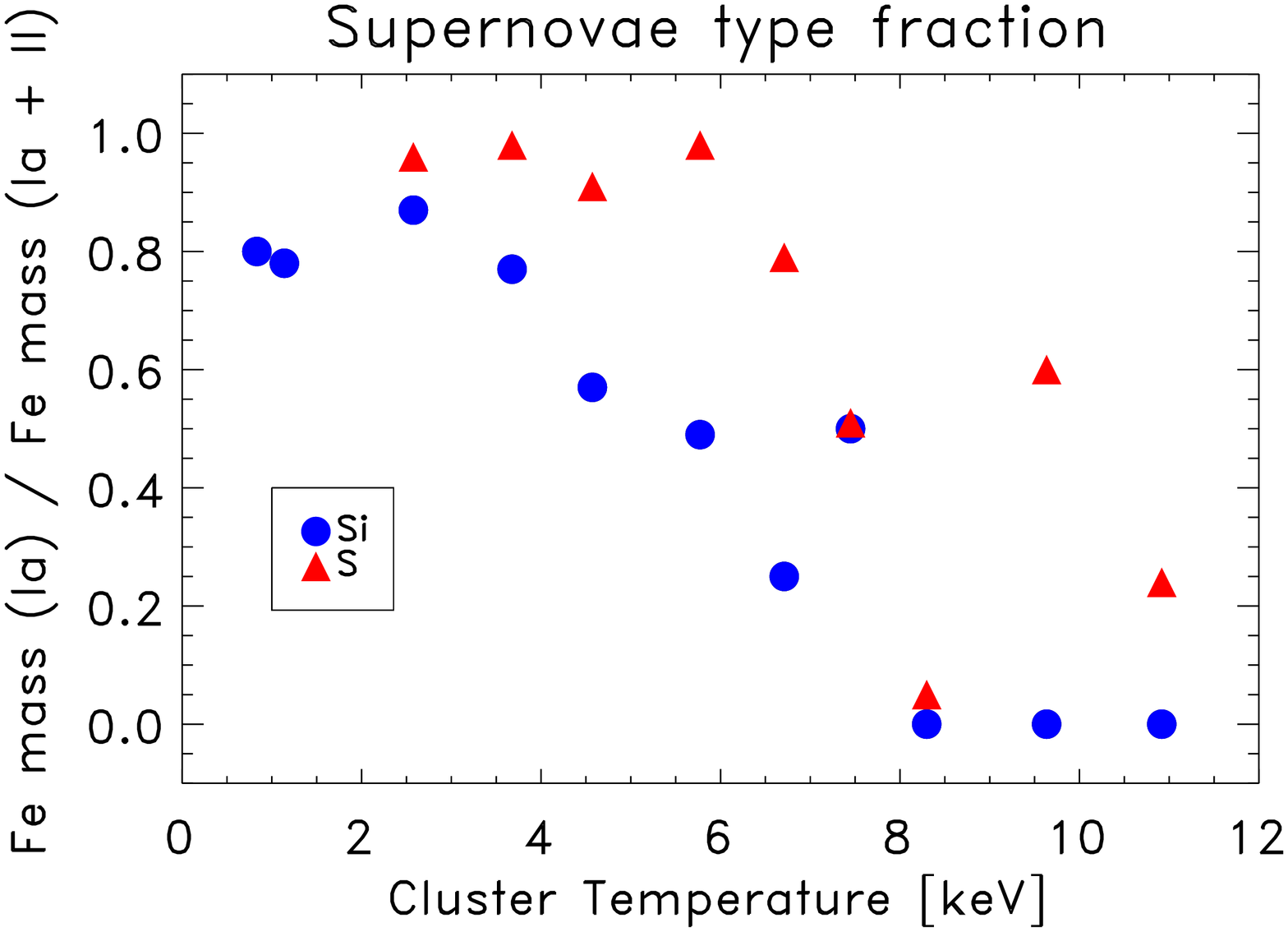}}}
\end{center}
\caption{The supernovae-type ratio derived from the silicon and sulfur
abundances.  The \sno\ yields of the W7 model \citep{n97} and the
\snt\ yields of the IMF mass-averaged TNH-40 \citep{tnh40} model were
used to compute the Si/Fe and S/Fe abundance ratios for 100 mixtures
ranging from pure \sno\ enrichment to pure \snt\ enrichment.  The
measured data was compared to the model output in order to determine
the relative ratio of \sno\ and \snt.  A value of 1 on the plot
indicates enrichment by solely \sno\, while a value of 0 indicates
enrichment solely by \snt.
\label{sn_ratios_fig}}
\end{figure}

We used the data from the models to produce a table with the yields
and abundance ratios for 100 different mixtures of \sno\ and \snt\
ranging from pure \sno\ enrichment to pure \snt\ enrichment.
We then compare our two best-measured cluster abundance
ratios (Si/Fe and S/Fe) with the model ratios in order to arrive at a
SN type fraction.  The results are plotted in
Figure~\ref{sn_ratios_fig}
and show the SN type fraction derived from both Si/Fe and S/Fe as a
function of cluster temperature.

\begin{figure*}
\begin{center}
\resizebox{!}{8in}{\rotatebox{0}{\includegraphics{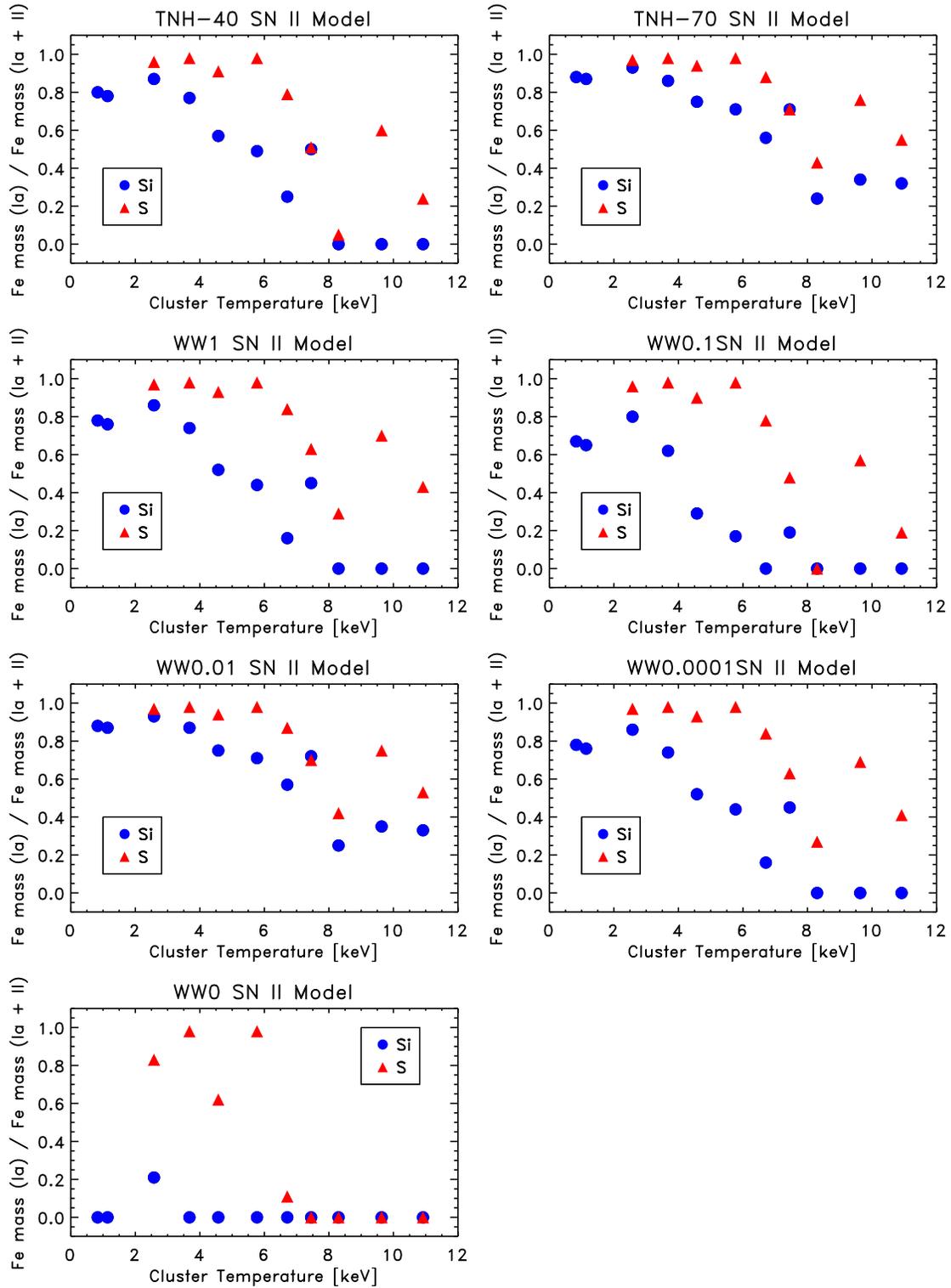}}}
\end{center}
\caption{The supernovae-type ratio derived from the silicon and sulfur
abundances, comparing the results from many different \snt\
models. This figure is similar to Figure~\ref{sn_ratios_fig}, but shows
that many different \snt\ models all have difficulty producing a
consistent type fraction from the silicon and sulfur data.  Also, the
trend of cooler clusters predominantly enriched by type \sno\ products
and hotter clusters enriched by \snt\ products does not depend on the
\snt\ model used.  Again, a value of 1 on the plot indicates
enrichment by solely \sno\, while a value of 0 indicates enrichment
solely by \snt.
\label{model_comparison}}
\end{figure*}

The difficulty of calculating SN yields leads to acknowledged
uncertainties for some of the elements of about a factor of two
\citep{glm97}. In order to investigate the effects of different SN
models on the derived SN fraction, we have compared the SN fractions
derived from several different SN models in
Figure~\ref{model_comparison}.
Each SN fraction determination uses the W7 \sno\ model as revised in
\cite{n97}, but uses a different \snt\ model.  The result of this
comparison indicates that different \snt\ models can change the
average value for the derived SN fraction, but does not change the
fact that none of the seven \snt\ models results in a consistent SN
fraction from both the cluster Si/Fe and S/Fe data.

The results we have found for the cluster intermediate element
abundances do not fit the standard model where most of the elements
are produced in a simple mix of standard \sno\ and \snt\ events.  The
observed silicon abundance is too high with respect to iron, and the
observed sulfur abundance too low.  Calcium and argon are also too
low.  The SN type fractions derived from the two abundance ratios are
not consistent with each other, and change dramatically as a function
of cluster temperature.  \textit{These results indicate that a unique,
consistent decomposition of the ICM enrichment into \sno\ and \snt\
contributions is not possible}.

Faced with these inconsistencies, new models must be examined
that can better explain our results.  Some possibilities are included
in the following section.

\subsection{Expected Abundances from Standard SN Models}

\begin{deluxetable*}{cccccccc}
\tabletypesize{\small}
\tablecaption{Observed and Model Abundances\label{model_abuns}}\tablewidth{0pt}
\tablehead{
\colhead{Element} & 
\colhead{\asca\ } & 
\colhead{\snt\ } & 
\colhead{\sno\ } &
\colhead{\snt+\sno\ } & 
\colhead{+\sntx\tablenotemark{a}} & 
\colhead{+\sntx\tablenotemark{b}} & 
\colhead{+\sntx\tablenotemark{c}}
}

\renewcommand{\arraystretch}{1.0}
\startdata
Si	& 0.6--0.8	& 0.34	& 0.095		& 0.4\phn	& 0.7\phn	& 0.7	& 0.7\phn\\
S	& 0.1--0.4	& 0.20	& 0.09\phn	& 0.3\phn	& 0.4\phn	& 0.3	& 0.4\phn\\
Ar	& 0.0--0.2	& 0.17	& 0.075		& 0.25		& 0.3\phn	& 0.2	& 0.3\phn\\
Ca	& 0.0--0.2	& 0.19	& 0.09\phn	& 0.3\phn	& 0.35		& 0.3	& 0.35\\
Fe	& 0.4		& 0.14	& 0.26\phn	& 0.4\phn	& 0.4\phn	& 0.4	& 0.4\phn\\
Ni	& 0.8--1.5	& 0.16	& 0.85\phn	& 1.0\phn	& 1.0\phn	& 1.0	& 1.0\phn
\enddata
\tablecomments{Abundances are given with respect to the solar values
in Table~\ref{solar_abun}, column~3.}
\tablenotetext{a}{70 \msun\ progenitors \citep{tnh96} in addition to \snt\ and \sno.}
\tablenotetext{b}{0.01 solar abundance, $>$ 30 \msun\ progenitors
\citep{ww95} in
addition to \snt\ and \sno.}
\tablenotetext{c}{70 \msun\ He core Pop III progenitors \citep{hw02}
in addition to \snt\ and \sno.}
\end{deluxetable*}

There are many different supernovae yield computations in the
scientific literature.  For \snt, there are variations in the
progenitor masses, metallicities and internal structure; in explosion
energy and placement of the mass cut; and in adopted nuclear reaction
rates.  For \sno, parameters include the central white dwarf
progenitor density and burning front propagation speed.  Initially, we
consider the standard W7 deflagration \sno\ model and
Salpeter-IMF-averaged \snt\ yields from \cite{n97}.  In all cases, a
combination of \sno\ and \snt\ yields is necessary to match the
observed results.

About 0.02 \snt\ events per current solar mass of stars will occur if
most of the stars in cluster galaxies form over an interval much
shorter than the age of the universe (indicated by the predominance of
early-type galaxies in clusters) and if star formation proceeds with a
standard IMF \citep{k02}.  This calculation takes the mass lost to
winds and remnants into account and assumes the \snt\ progenitors are
stars with original mass $>$ 8 \msun.  Column~2 of
Table~\ref{model_abuns} shows the observational range we aim to
explain, Column~3 of Table~\ref{model_abuns} shows the model
abundances from ICM enrichment by \snt\ (for the gas-to-star mass
ratio of 10 that is typical of a rich cluster).

If the assumption is made that \sno\ provide the remainder of the iron
(The enrichment levels due solely to \sno\ enrichment are in column~4 of
Table~\ref{model_abuns}), then the total expected abundances (\snt\ +
\sno) are as shown in column~5 of Table~\ref{model_abuns}.  The number of
\sno\ required corresponds to an average of 1.8 SNU over $10^{10}$
years for a typical ratio of ICM mass to stellar blue light of 40,
significantly higher than the estimated optical rate at $z=0$
\citep{c97}.  This combination of \snt\ and \sno\ accounts for the
observed abundances of nickel and sulfur (compare columns~2 and 5 of
Table~\ref{model_abuns}).  However, only about half of the observed
silicon is accounted for and calcium and argon are slightly
overproduced.  Different models must be investigated to make up this
significant shortfall.  The most variation in SN yields is in \snt\
models, and we explore some of these alternatives below.

Unfortunately, many \snt\ models, in particular those of \cite{ww95}
and \cite{r02}, generally derive higher yields of sulfur, argon and
calcium that exacerbate the conflict between theory and observations.
In addition, \snt\ iron yields are uncertain by at least a factor of
two \citep{glm97} because of their sensitivity to the assumed
mass cut.  If one arbitrarily and exclusively increases the \snt\
contribution to the iron yield by a factor of two, then the \sno\
contribution to sulfur, argon and calcium goes down by a factor of two
and is in marginal agreement with the observations.  Unfortunately,
this scenario increases the silicon deficit and the nickel abundance
falls to the unacceptably low value of about half solar.

Models that explicitly synthesize more iron (\textit{e.g.}, those of
\cite{ww95} with enhanced explosion energy) also generally produce
more sulfur, argon and calcium that enhances the problem with those
elements.  However, the \cite{ww95} enhanced energy models having zero
metallicity progenitors have a very different nucleosynthetic profile
--- the production of iron and nickel averaged over the IMF is doubled
without an increase in sulfur, argon and calcium.  Unfortunately,
silicon and nickel remain underproduced by a factor of two if these
\snt\ yields are adopted.

Other variations on the \snt\ yields also fail to explain the low
ratio of sulfur, argon and calcium to silicon.  Neither adopting
delayed detonation \sno\ models \citep{f02} nor increasing the silicon
abundance by assuming a flat IMF \citep{lm96} solves the problem.
\citet{f02} explain the abundance pattern and its radial variation in
M87 by proposing (1) a radially increasing SNII/SNIa ratio, (2) high
Si and S yields from SNIa (favoring delayed detonation models) and an
{\it ad hoc} reduction in SNII S yields, and (3) a radial variation in
SNIa yields (corresponding to delayed detonation models with different
deflagration-to-detonation transition densities). While this scenario
appears promising with regard to the galaxy-dominated inner regions of
rich clusters and, perhaps groups, there are difficulties -- as
\citet{f02} acknowledge -- in explaining the pattern in the large
scale abundances of high-temperature clusters.

One possible solution is to assume a flat IMF to boost \snt\ silicon
production (and account for at least half of the iron), combined with
\textit{ad hoc} reductions in sulfur, argon and calcium yields by a
factor of two.  An increase in the \snt\ nickel yields (uncertain by a
factor of two because of mass cut and core neutron excess
uncertainties, \citealt{n99}) would also have to be instated to
account for the lowered nickel from the reduced \sno\ output.  On the
other hand, if a standard IMF is maintained, then less significant
decreases in \snt\ production of sulfur, argon and calcium are needed.
For this case, the silicon deficit would have to be made up by another
mechanism; enhanced hydrostatic production of silicon or a suppressed
efficiency of silicon burning are possibilities.

\subsection{Alternate Enrichment Scenarios}
\label{3types}

If the observed cluster iron abundance and stellar elemental
abundances are used to set the contribution from supernovae, then the
theoretical yields will meet or slightly exceed the \asca\ abundance
observations for argon and calcium while silicon itself will be
significantly underproduced by a factor of $\sim$2.  If we accept the
standard \snt\ enrichment given in column~3 of
Table~\ref{model_abuns}, then we must have an additional
nucleosynthetic source that produces substantial silicon, but little
or no sulfur, argon, or calcium.  Some possibilities exist in the
literature, and we explore some of these scenarios below.  We have
chosen single mass \snt\ models that can selectively enhance silicon,
and collectively call these models ``\sntx'' to distinguish then from
the canonical \snt\ models.  For the models presented below, we
calculate the number of \sntx\ events necessary to produce the
observed amounts of iron and silicon that are not produced by regular
\snt.

\subsubsection{Very Massive Stars}

\cite{tnh96} have calculated the yields for a 70 M$_\sun$ progenitor
with solar metallicity.  The abundance ratios in the ejecta with respect
to silicon are: 0.4, 0.28, 0.26, 0.048, and 0.12 relative to solar for
S, Ar, Ca, Fe, and Ni.  With these results, the observational
overabundance of silicon can be accounted for with $\sim 2 \times
10^{-4}$ \sntx\ per solar mass in the ICM.  This rate will provide the
extra silicon necessary, is only about 10\% of the total number of
\snt\ expected by integrating a normal IMF, and only produces minor
perturbations in the other elements.  The abundances expected from the
combination of \sno, canonical \snt, and \sntx\ from very massive
progenitors is given in column~6 of Table~\ref{model_abuns}.

\subsubsection{Massive, Metal-poor Stars}

Very small amounts of the elements heavier than silicon are produced
in the high mass ($>$ 25 M$_\sun$), low-metallicity (0.01, 0.1 solar)
models of \cite{ww95}.  These yields give less than 10\% of the solar
ratios (with respect to silicon) for elements heavier than silicon.
Hence, silicon abundances can be increased without causing the other
elements to exceed their observed values.  In order to produce the
observed abundances, about $5.7 \times 10^{-4}$ \sntx\ per solar mass
in the ICM are necessary, assuming progenitors of 30--40 M$_\sun$ with
0.01 solar abundance and an IMF that goes as $M^{-1}$.  The abundances
expected from the combination of \sno, canonical \snt, and \sntx\ from
massive, metal-poor progenitors is given in column~7 of
Table~\ref{model_abuns}.

\subsubsection{Population III Stars}

The ejecta of supernovae preceded by very massive, zero metallicity
progenitor stars (Pop~III stars) have a substantially different
composition than standard \snt\ ejecta (\textit{e.g.}, \cite{hw02}).  If the
Pop~III mass function is dominated by the low mass (He cores $< 80$
\msun) models of \cite{hw02}, then the relative Si overabundance from
observations can be explained.  If higher core masses are used, the
silicon overabundances are not as well explained, but more modest
underabundances are found for sulfur, argon, and calcium.  For
example, about $2.3 \times 10^{-5}$ \sntx\ events per solar mass in
the ICM from progenitors with 70 \msun\ He cores matches our observed
abundances and produce the abundances given in column~8 of
Table~\ref{model_abuns}.

\begin{figure}[!t]
\begin{center}
\resizebox{\columnwidth}{!}{\rotatebox{90}{\includegraphics{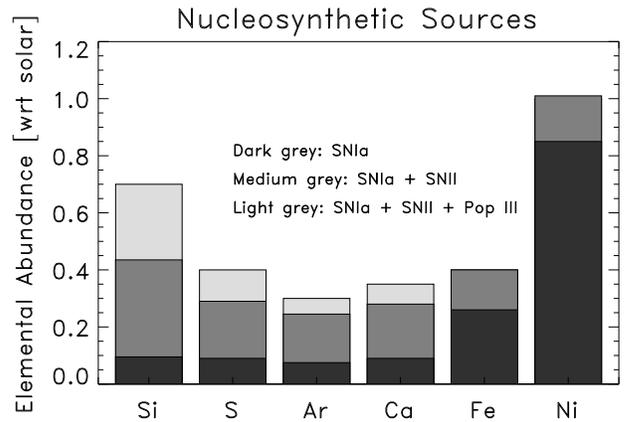}}}
\end{center}
\caption{A decomposition of the intermediate element abundances into
their nucleosynthetic origins based on Table~\ref{model_abuns}.  The
lowest band in each bar is the contribution from \sno, the middle band
adds the yields from the canonical \snt, and the top band includes the
contribution from an early population of high mass, low metallicity Pop
III supernovae progenitors.  Iron and nickel both are dominated by
\sno\ products, with no contribution by \sntx\ from Pop~III
progenitors. 
\label{decomposition}}
\end{figure}

Figure~\ref{decomposition} 
is based on Table~\ref{model_abuns} and shows graphically what
proportion of each element's abundance comes from \sno, \snt, and
\sntx.  The three different \sntx\ models all produce similar results,
and provide the silicon necessary to match the observational
overabundance without unduly exacerbating the low observed abundances
of sulfur, argon and calcium.

\subsection{Discussion}

The anomalous abundance patterns found here in galaxy clusters are
unique.  The relative simplicity and uncomplicated nature of the X-ray
emission from clusters and their role as a repository for all the
enriched gas produced by supernovae makes them important objects for
understanding the production and evolution of metals in the universe.

However, no combination of \sno\ and \snt\ products using current
theoretical yields can produce the abundances observed with
\asca. Manipulating the standard models by changing the IMF or
appealing to different physical models such as delayed detonation can
help mitigate only some of the inconsistencies between models and
observations.  A new source of metals is needed in order
satisfactorily explain the cluster abundances.  The low ratios of
S/Si, Ar/Si and Ca/Si are indicative of a fundamentally different mode
of heavy metal enrichment.  We have identified three different SN
models that can produce these metals in the correct proportions and
have calculated the number of events necessary to match the
observations when combined with metals produced in ordinary \sno\ and
\snt.  These models all have in common very massive, metal poor
progenitor stars.

It is natural to associate these with the earliest generation of
(Pop~III) stars.  This primordial population must have existed, and
its distinct characteristics are well matched to those required to
explain the ICM abundance anomalies: zero metallicity, and a top-heavy
IMF that has an important contribution from very massive stars
\citep{bcl99, abn00}.  Enhanced explosion energies are indicated, and
nucleosynthetic production is high with an abundance pattern quite
distinct from other \snt\ \citep{hw02}.  

\cite{l01} proposed hypernovae associated with Pop~III as a means of
enhancing silicon relative to oxygen in the ICM, and inferred a
hypernovae rate of the same order as the \sntx\ from Pop~III stars
derived here.  The precise required number is subject to assumed
values of the ICM/stellar ratio, the IMF and \snt\ yields for Pop~II
stars, and the IMF and \sntx\ yields for Pop~III stars, but is in the
range of 10--30 times larger than the average value ($\Omega_{III}
\sim 4 \times 10^{-6}$) predicted in the semi-analytic models of
\cite{og96}.  This could be the result of enhanced primordial star
formation in these extremely overdense environments.  Evidence also
exists for more nearly solar abundance ratios in less massive clusters
\citep{f02}.

Both observations and theory suggest the idea that a large population
of massive, metal-poor Pop~III stars was present at very high redshift
and constituted the first generation of stars.  The heavy metal
products from this first generation would have been formed at the same
time as the first galaxies or even before, and would now be widely
dispersed in the ICM.  Galaxy clusters serve as the largest
repositories of both enriched gas from cluster galaxies and of gas
expelled by the earliest generations of stars and then accreted onto
clusters.  A combination of X-ray observations well matched to
observing metal abundances in clusters and the importance of galaxy
clusters as large retainers of Pop~III enriched gas make these
observations one of the best views onto the earliest generations of
stars in the universe.

\section{SUMMARY}

We have presented intermediate element abundances for galaxy clusters
based on \asca\ observations.  Our measurements of the iron and
silicon abundances agree with the past \asca\ results of \cite{fuk97}
and \cite{ml96}, but achieve much higher precision and extend the
temperature range from 0.5--12\,keV.  The measurements of the
individual element abundances show some surprising new results: silicon
and sulfur do not track each other as a function of temperature in
clusters, and argon and calcium have much lower abundances than
expected.  

These results show that the $\alpha$-elements do not behave
homogeneously as a single group.  The unexpected abundance trends with
temperature probably indicate that different enrichment mechanisms are
important in clusters with different masses.  The wide scatter in
$\alpha$-element abundances at a single temperature could indicate
that SN models need some fine tuning of the individual element yields,
or that a different population of SN needs to be considered as
important to metal enrichment in clusters.

We have also attempted to use our measured abundances to constrain the
SN types that caused the metal enrichment in clusters.  A first
attempt to split the metal content into contributions from canonical
\sno\ and \snt\ led to inconsistent results with both the individual
elemental abundances and with the abundance ratios of our most well
measured elements.  An investigation of different SN models also could
not lead to a scenario consistent with the measured data, and we
deduce that no combination of \sno\ and \snt\ fits the data.  Another
source of metals is needed.

This extra source of metals must be able to produce enough silicon to
match the measurements, but not so much sulfur, argon, or calcium to
exceed them.  An investigation of SN models in the literature led to
three separate models that could fulfill these requirements.  All
three models were similar, and had as progenitor stars massive and/or
metal poor stars.  The combination of canonical \sno\ and \snt\ with
one of the new models does a much better job of matching the
observations.  These sorts of massive, metal poor progenitor stars are
exactly the stars that are supposed to make up the very early
Population III stars.  The conjunction of our required extra source of
metals with SN from population III-like progenitors supports the idea
that a significant amount of metal enrichment was from the very
earliest stars.

Clusters are a unique environment for elemental abundance measurements
because they retain all the metals produced in them.  The relatively
uncomplicated physical environment in clusters also allows
well-understood abundance measurements.  Future abundance analyses
using a large sample of \xmm\ data will allow an even better
understanding of the SN types and enrichment mechanisms important in
galaxy clusters.  

\acknowledgements

The authors would like to thank K. Gendreau for early assistance with
this work and K. Kuntz for continued useful conversations and
technical advice.  The prompt and useful comments of the referee,
A. Finoguenov, were also greatly appreciated.  This research has made
extensive use of data obtained from the High Energy Astrophysics
Science Archive Research Center (HEASARC), provided by NASA's Goddard
Space Flight Center; the NASA/IPAC Extragalactic Database (NED) which
is operated by the Jet Propulsion Laboratory, California Institute of
Technology, under contract with the National Aeronautics and Space
Administration; and the SIMBAD database, operated at CDS, Strasbourg,
France.  The authors extend their sincere thanks to the people
responsible for making these resources available online.



\begin{thebibliography}{}

\bibitem[Abel, Bryan, \& Norman(2000)]{abn00} Abel, T., 
Bryan, G.~L., \& Norman, M.~L.\ 2000, \apj, 540, 39 
\bibitem[Allende Prieto, Lambert, \& Asplund(2001)]{apla01} Allende
Prieto, C., Lambert, D.~L., \& Asplund, M.\ 2001, \apjl, 556, L63
\bibitem[Anders \& Grevesse(1989)]{angr89} Anders, E.~\& Grevesse, N.\
1989, \gca, 53, 197
\bibitem[Angelini, Loewenstein, \& Mushotzky(2001)]{angelini01}
Angelini, L., Loewenstein, M., \& Mushotzky, R.~F.\ 2001, \apjl, 557,
L35
\bibitem[Arnaud et al.(2001)]{arnaud01} Arnaud, M., Neumann, D.~M.,
Aghanim, N., Gastaud, R., Majerowicz, S., \& Hughes, J.~P.\ 2001,
\aap, 365, L80
\bibitem[Arnaud et al.(1992)]{a92}Arnaud, M., Rothenflug, R., Boulade,
O., Vigroux, L., \& Vangioni-Flam, E.\ 1992, \aap, 254, 49
\bibitem[Becker et al.(1979)]{b79} Becker, R.~H., Smith, B.~W., White,
N.~E., Holt, S.~S., Boldt, E.~A., Mushotzky, R.~F., \& Serlemitsos,
P.~J.\ 1979, \apjl, 234, L73
\bibitem[Bell, McIntosh, Katz, \& Weinberg(2003)]{bell03} Bell, E.~F.,
McIntosh, D.~H., Katz, N., \& Weinberg, M.~D.\ 2003, ArXiv
Astrophysics e-prints, 2543
\bibitem[B{\" o}hringer et al.(2001)]{bohr01} B{\" o}hringer, H.~et
al.\ 2001, \aap, 365, L181
\bibitem[Bromm, Coppi, \& Larson(1999)]{bcl99} Bromm, V., 
Coppi, P.~S., \& Larson, R.~B.\ 1999, \apjl, 527, L5 
\bibitem[Cappellaro et al.(1997)]{c97} Cappellaro, E., Turatto, M.,
Tsvetkov, D.~Y., Bartunov, O.~S., Pollas, C., Evans, R., \& Hamuy, M.\
1997, \aap, 322, 431
\bibitem[DeGrandi(2003)]{dig} DeGrandi, S. 2003 in proceedings of the
Ringberg Cluster Conference
\bibitem[Dupke \& Arnaud(2001)]{da} Dupke, R.~A.~\&
Arnaud, K.~A.\ 2001, \apj, 548, 141
\bibitem[Dupke \& White(2000a)]{dwa} Dupke, R.~A.~\&
White, R.~E.\ 2000a, \apj, 528, 139
\bibitem[Dupke \& White(2000b)]{dwb} Dupke, R.~A.~\&
White, R.~E.\ 2000b, \apj, 537, 123
\bibitem[Finoguenov, Arnaud, \& David(2001)]{f01} 
Finoguenov, A., Arnaud, M., \& David, L.~P.\ 2001, \apj, 555, 191 
\bibitem[Finoguenov, David, \& Ponman(2000)]{f00} 
Finoguenov, A., David, L.~P., \& Ponman, T.~J.\ 2000, \apj, 544, 188 
\bibitem[Finoguenov et al.(2002)]{f02} Finoguenov, A., Matsushita, K.,
B{\" o}hringer, H., Ikebe, Y., \& Arnaud, M.\ 2002, \aap, 381, 21
\bibitem[Fukazawa(1997)]{fuk97}Fukazawa, Y. 1997, Ph.D. Dissertation,
University of Tokyo
\bibitem[Fukazawa et al.(1998)]{fuk98} Fukazawa, Y., Makishima, K.,
Tamura, T., Ezawa, H., Xu, H., Ikebe, Y., Kikuchi, K., \& Ohashi, T.\
1998, \pasj, 50, 187
\bibitem[Gastaldello \& Molendi(2002)]{gm02}
Gastaldello, F.~\& Molendi, S.\ 2002, \apj, 572, 160
\bibitem[Gibson, Loewenstein, \& Mushotzky(1997)]{glm97} Gibson,
B.~K., Loewenstein, M., \& Mushotzky, R.~F.\ 1997, \mnras, 290, 623
\bibitem[Heger \& Woosley(2002)]{hw02} Heger, A.~\& Woosley, S.~E.\
2002, \apj, 567, 532
[\bibitem[Henry \& Worthey(1999)]{hw99}Henry, R.~B.~C., \& Worthey,
G. 1999, PASP, 111, 119
\bibitem[Horner(2001)]{horner01}Horner, D. 2001, Ph.D. Dissertation,
Department of Astronomy, University of Maryland College Park
\bibitem[Horner et al.(ApJS submitted)]{horner03} Horner, D.~J.,
Baumgartner, W.~H., Gendreau, K.~C., \& Mushotzky, R.~F. 2003, \apjs,
submitted
\bibitem[Hwang et al.(1999)]{hwang99} Hwang, U., Mushotzky, R.~F.,
Burns, J.~O., Fukazawa, Y., \& White, R.~A.\ 1999, \apj, 516, 604
\bibitem[Irwin, Athey, \& Bregman(2003)]{irwin03} Irwin, J.~A., Athey,
A.~E., \& Bregman, J.~N.\ 2003, \apj, 587, 356 \bibitem[Ishimaru \&
Arimoto(1997)]{ia97} Ishimaru, Y.~\& Arimoto, N.\ 1997, \pasj, 49, 1
\bibitem[Kroupa(2002)]{k02} Kroupa, P. 2002, Science, 296, 82
\bibitem[Grevesse \& Sauval(1998)]{grsa98} Grevesse, N.~\& Sauval,
A.~J.\ 1998, Space Science Reviews, 85, 161
\bibitem[Grevesse \& Sauval(1999)]{grsa99} Grevesse, N.~\& Sauval,
A.~J.\ 1999, \aap, 347, 348
\bibitem[Guainazzi \& Molendi(1999)]{gm99} Guainazzi, M.~\& 
Molendi, S.\ 1999, \aap, 351, L19
\bibitem[Loewenstein(2001)]{l01} Loewenstein, M.\ 2001, 
\apj, 557, 573 
\bibitem[Loewenstein \& Mushotzky(1996)]{lm96} Loewenstein, M.~\&
Mushotzky, R.~F.\ 1996, \apj, 466, 695
\bibitem[Matsumoto et al.(1996)]{matsu96} Matsumoto, H., 
Koyama, K., Awaki, H., Tomida, H., Tsuru, T., Mushotzky, R., \& Hatsukade, 
I.\ 1996, \pasj, 48, 201 
\bibitem[Matsushita, Finoguenov, \& B{\"o}hringer(2003)]{mat03}
Matsushita, K., Finoguenov, A., \& B{\" o}hringer, H.\ 2003, \aap,
401, 443
\bibitem[Mitchell, Culhane, Davison, \& Ives(1976)]{mitch76} Mitchell,
R.~J., Culhane, J.~L., Davison, P.~J.~N., \& Ives, J.~C.\ 1976,
\mnras, 175, 29P
\bibitem[Molendi \& Gastaldello(2001)]{mg01} Molendi, S.~\& 
Gastaldello, F.\ 2001, \aap, 375, L14 
\bibitem[Mori et al.(2000)]{mori} Mori, K.~et al.\ 2000, \procspie,
4012, 539
\bibitem[Mushotzky(1983)]{mush83} Mushotzky, R.~F.\ 1983, Presented at
the Workshop on Hot Astrophys.~Plasmas, Nice, 8-10 Nov.~1982, 83,
33826
\bibitem[Mushotzky et al.(1981)]{m81} Mushotzky, R.~F., Holt, S.~S.,
Boldt, E.~A., Serlemitsos, P.~J., \& Smith, B.~W.\ 1981, \apjl, 244,
L47
\bibitem[Mushotzky \& Loewenstein(1997)]{ml97} Mushotzky, 
R.~F.~\& Loewenstein, M.\ 1997, \apjl, 481, L63 
\bibitem[Mushotzky et al.(1996)]{ml96} Mushotzky, R., Loewenstein, M.,
Arnaud, K.~A., Tamura, T., Fukazawa, Y., Matsushita, K., Kikuchi, K.,
\& Hatsukade, I.\ 1996, \apj, 466, 686
\bibitem[Mushotzky et al.(1978)]{mush78} Mushotzky, R.~F.,
Serlemitsos, P.~J., Boldt, E.~A., Holt, S.~S., \& Smith, B.~W.\ 1978,
\apj, 225, 21
\bibitem[Nakamura et al.(1999)]{n99} Nakamura, T., Umeda, H., Nomoto,
K., Thielemann, F., \& Burrows, A.\ 1999, \apj, 517, 193
\bibitem[Nomoto et al.(1997a)]{tnh40} Nomoto, K., Hashimoto, M.,
Tsujimoto, T., Thielemann, F.-K., Kishimoto, N., Kubo, Y., \&
Nakasato, N.\ 1997, Nuclear Physics A, 616, 79
\bibitem[Nomoto et al.(1997b)]{n97} Nomoto, K., Iwamoto, K., Nakasato,
N., Hashimoto, Thielemann, F.-K., Brachwitz, F., Tsujimoto, T., Kubo,
Y., \& Kishimoto, N.\ 1997, Nuclear Physics A, 621, 467 
\bibitem[Ostriker \& Gnedin(1996)]{og96} Ostriker, J.~P.~\& 
Gnedin, N.~Y.\ 1996, \apjl, 472, L63 
\bibitem[Peimbert, Carigi, \& Peimbert(2001)]{peimbert01} Peimbert,
M., Carigi, L., \& Peimbert, A.\ 2001, Astrophysics and Space Science
Supplement, 277, 147
\bibitem[Peterson et al.(2002)]{peterson02} Peterson, J.~R.,
Kahn, S.~M., Paerels, F.~B.~S., Kaastra, J.~S., Tamura, T., Bleeker,
J.~A.~M., Ferrigno, C., \& Jernigan, J.~G.\ 2002, ArXiv Astrophysics
e-prints, 10662
\bibitem[Pettini et al.(2002)]{pettini02} Pettini, M., Rix, S.~A.,
Steidel, C.~C., Adelberger, K.~L., Hunt, M.~P., \& Shapley, A.~E.\
2002, \apj, 569, 742
\bibitem[Pipino, Matteucci, Borgani, \& Biviano(2002)]{pipino02} 
Pipino, A., Matteucci, F., Borgani, S., \& Biviano, A.\ 2002, New 
Astronomy, 7, 227 
\bibitem[Pompeia, Barbuy, \& Grenon(2003)]{pompeia03}Pompeia, L.,
Barbuy, B., \& Grenon, M.\ 2003, ArXiv Astrophysics e-prints, 04282
\bibitem[Prochaska et al.(2000)]{prochaska00} Prochaska,
J.~X., Naumov, S.~O., Carney, B.~W., McWilliam, A., \& Wolfe, A.~M.\
2000, \aj, 120, 2513 
\bibitem[Prochaska \& Wolfe(2002)]{pw02} Prochaska, J.~X.~\& 
Wolfe, A.~M.\ 2002, \apj, 566, 68 
\bibitem[Rauscher, Heger, Hoffman, \& Woosley(2002)]{r02} Rauscher,
T., Heger, A., Hoffman, R.~D., \& Woosley, S.~E.\ 2002, \apj, 576, 323
\bibitem[Rothenflug, Vigroux, Mushotzky, \& Holt(1984)]{r84}
Rothenflug, R., Vigroux, L., Mushotzky, R.~F., \& Holt, S.~S.\ 1984,
\apj, 279, 53
\bibitem[Sakelliou et al.(2002)]{sak02} Sakelliou, I.~et
al.\ 2002, \aap, 391, 903
\bibitem[Serlemitsos et al.(1977)]{s77} Serlemitsos, P.~J., Smith,
B.~W., Boldt, E.~A., Holt, S.~S., \& Swank, J.~H.\ 1977, \apjl, 211,
L63
\bibitem[Smith et al.(2001)]{smith01}Smith, R. K., Brickhouse,
N. S., Liedahl, D. A., \& Raymond, J. S. 2001, ApJ, 556, 91
\bibitem[Songaila(2001)]{song01} Songaila, A.\ 2001, \apjl, 561, L153
\bibitem[Songaila \& Cowie(2002)]{song02} Songaila, A.~\& Cowie,
L.~L.\ 2002, \aj, 123, 2183
\bibitem[Stasinska(2002)]{stas02} Stasinska, G.\ 2002, ArXiv
Astrophysics e-prints, 7500
\bibitem[Tamura et al.(2001)]{tamura01} Tamura, T., Bleeker, J.~A.~M.,
Kaastra, J.~S., Ferrigno, C., \& Molendi, S.\ 2001, \aap, 379, 107
\bibitem[Tanaka, Inoue, \& Holt(1994)]{tih}Tanaka, Y., Inoue, H.,
\& Holt, S. S. 1994, \pasj, 46, L37
\bibitem[Thielemann, Nomoto, \& Hashimoto(1996)]{tnh96} Thielemann,
F., Nomoto, K., \& Hashimoto, M.\ 1996, \apj, 460, 408
\bibitem[Timmes, Woosley, \& Weaver(1995)]{tww95} Timmes, F.~X.,
Woosley, S.~E., \& Weaver, T.~A.\ 1995, \apjs, 98, 617
\bibitem[Wilms, Allen, \& McCray(2000)]{wam}Wilms, J., Allen, A.,
\& McCray, R. 2000, \apj, 542, 914
\bibitem[Woosley \& Weaver(1995)]{ww95} Woosley, S.~E.~\& 
Weaver, T.~A.\ 1995, \apjs, 101, 181 
\bibitem[Yaqoob et al.(2000)]{yaqoob00} Yaqoob, T. and the \asca\
team. ASCA GOF Calibration Memo (ASCA-CAL-00-06-01, v1.0
06/05/00)
\end{thebibliography}
\end{document}